\documentclass[useAMS,usenatbib]{mnras}

\usepackage{hyperref}
\usepackage{graphicx}
\usepackage{verbatim}
\usepackage{pdflscape}

\title[LAMBDAR G10 photometry]{G10/COSMOS:  38 band (far-UV to far-IR) panchromatic photometry using LAMBDAR}
\author[Andrews et al.]{S. K. Andrews$^{1}$\thanks{E-mail:stephen.andrews@icrar.org}, S. P. Driver$^{1, 2}$, L. J. M. Davies$^{1}$,  
\newauthor Prajwal R. Kafle$^{1}$, Aaron S. G. Robotham$^{1}$, Angus H. Wright$^{1}$
\footnotemark[1]\thanks{Herschel is an ESA space observatory with science instruments provided by European-led Principal Investigator consortia and with important participation from NASA.}\\
$^{1}$International Centre for Radio Astronomy Research, The University of Western Australia, 35 Stirling Highway, Crawley, WA 6009, Australia \\
$^{2}$School of Physics \& Astronomy, University of St Andrews, North Haugh, St Andrews, KY16 9SS, United Kingdom }
\begin{document}

\date{Accepted 1988 December 15. Received 1988 December 14; in original form 1988 October 11}

\pagerange{\pageref{firstpage}--\pageref{lastpage}} \pubyear{2002}

\maketitle

\label{firstpage}

\begin{abstract}
We present a consistent total flux catalogue for a $\sim$1 deg$^2$ subset of the COSMOS region (R.A. $\in [149.55\degr, 150.65\degr]$, DEC $\in [1.80\degr, 2.73\degr]$) with near-complete coverage in 38 bands from the far-ultraviolet to the far-infrared. We produce aperture matched photometry for 128,304 objects with $i < 24.5$ in a manner that is equivalent to the \citet{wright16a} catalogue from the low-redshift ($z < 0.4$) Galaxy and Mass Assembly (GAMA) survey. This catalogue is based on publicly available imaging from \textit{GALEX}, CFHT, Subaru, VISTA, \textit{Spitzer} and Herschel, contains a robust total flux measurement or upper limit for every object in every waveband and complements our re-reduction of publicly available spectra in the same region. We perform a number of consistency checks, demonstrating that our catalogue is comparable to existing data sets, including the recent COSMOS2015 catalogue \citep{laigle16}. We also release an updated \citet{davies15} spectroscopic catalogue that folds in new spectroscopic and photometric redshift data sets. The catalogues are available for download at \url{http://cutout.icrar.org/G10/dataRelease.php}. Our analysis is optimised for both panchromatic analysis over the full wavelength range and for direct comparison to GAMA, thus permitting measurements of galaxy evolution for $0 < z < 1$ while minimising the systematic error resulting from disparate data reduction methods.

\end{abstract}


\section{Introduction}
\label{sec:intro}

Wide-area multiwavelength surveys such as the Sloan Digital Sky Survey (SDSS; \citealt{york00}) and the UKIRT (UK Infrared Telescope) Deep Sky Survey \citep{lawrence07} have enabled the study of large, statistical samples of galaxies. However, such surveys are generally limited to low redshifts ($z ~< 0.3$), a single facility, and one region of the electromagnetic spectrum --- usually the ultraviolet, optical or near-infrared. To produce a comprehensive picture of galaxy evolution, one must observe galaxies over an extensive range of wavelengths to probe multiple physical properties. This requires the combination of multiple data sets across observatories and instruments, and thus the consolidation of disparate sensitivities, resolutions and data reduction techniques (see e.g. \citealt{driver15}).

Obtaining a consistent, optically-motivated photometric catalogue for large multiwavelength datasets is highly non-trivial \citep{laidler07,wright16a}. Naively position matching existing catalogues gives rise to the possibility of table mismatches, especially when joining high-resolution (resolution $\sim$ 0.8\arcsec) optical data to low-resolution far-infrared data (resolution $\sim$18\arcsec). Disparate data reduction methods, even though they may represent the most appropriate photometry in each individual band, may use differently sized and shaped apertures for the same object and hence probe different physical scales. More subtly, the different means of calculating errors by different survey teams will affect the quality of spectral energy distribution (SED) fits for a particular galaxy. \citet{wright16a} show that the use of a multiwavelength catalogue derived using the same data reduction procedure across the full wavelength range results in reduced photometric inconsistency, and improves the accuracy of SED fits and star formation rate estimators compared to an equivalent catalogue constructed from table matching alone. 

One technique to construct a consistent multiwavelength catalogue is a variation of (forced) matched aperture photometry, proceeding initially with aperture definition on a high-resolution image. The apertures are then propagated to the lower resolution data after convolution with the point spread function and appropriate deblending. Software packages implementing this technique include TFIT \citep{laidler07} and the Lambda-Adaptive MultiBand Deblending Algorithm in R (\textsc{lambdar}; \citealt{wright16a}).

One dataset that lends itself to the construction of such a catalogue is the Galaxy and Mass Assembly (GAMA; \citealt{driver11,liske15}) survey. GAMA is a highly complete low-redshift spectroscopic and multiwavelength imaging campaign that aims to characterise the distribution of energy, mass and structure from kiloparsec to megaparsec scales. The GAMA spectroscopic campaign targeted 230 degrees of sky using the AAOmega spectrograph on the 3.9~m Anglo-Australian Telescope, obtaining redshifts for $\sim$250,000 galaxies. This spectroscopy is complemented by ultra-violet imaging from the \textit{GALaxy Evolution eXplorer} (\textit{GALEX}; \citealt{martin05}), optical imaging from SDSS and the Kilo-Degree Survey (KiDS; \citealt{dejong15}), near-infrared imagery from the VISTA (Visible and Infrared Survey Telescope for Astronomy) Kilo-degree Infrared Galaxy (VIKING; \citealt{edge13}) survey, mid-infrared imagery from the \textit{Widefield Infrared Survey Explorer} \citep{wright10} and far-infrared imagery from \textit{Herschel}-Atlas \citep{eales10} --- see summary in \citet{driver15}. The project has examined a wide variety of topics, including the cosmic spectral energy distribution (e.g. \citealt{driver12}, Andrews et al. in prep), star formation rates \citep{davies16a}, large scale structure (e.g. \citealt{alpaslan14}), galaxy groups (e.g. \citealt{robotham11}), close pairs (e.g. \citealt{davies15b,davies16b}) and galaxy properties and structure (e.g. \citealt{taylor11,kelvin12,loveday15,moffett16}). The GAMA survey, while scientifically comprehensive, by design only probes the low redshift Universe ($z < 0.4$). It is therefore beneficial that an intermediate redshift ($0.3 < z < 1$) equivalent to GAMA is established in order to explore a broader time baseline.

The Cosmological Evolution Survey (COSMOS; \citealt{scoville07}) region, covering 2~deg$^2$ of sky centred on R.A. = 10h00m28.6s, DEC = +02$\degr$12\arcmin21\arcsec.0 is suitable for this purpose. The program is anchored by F814W observations of the field using the \textit{Hubble Space Telescope} and has been expanded to include deep observations spanning from X-ray wavelengths to radio -- with observations conducted and released using \textit{Chandra}, \textit{GALEX}, the Canada-France-Hawaii Telescope (CFHT), Subaru, VISTA, \textit{Spitzer} and \textit{Herschel}. Spectroscopic surveys in the COSMOS region include zCOSMOS \citep{lilly07,lilly09}, the PRIsm MUlti-object Survey (PRIMUS; \citealt{coil11,cool13}), the VIMOS-VLT Deep Survey (VVDS; \citealt{garilli08}), the VIMOS Ultra Deep spectroscopic survey \citep{lefevre15}, the FMOS-COSMOS survey \citep{silverman15}, 3D-\textit{HST} \citep{brammer12} and SDSS DR10 \citep{ahn14}, complemented by large catalogues of photometric redshifts \citep{ilbert09,muzzin14,laigle16}. COSMOS has been used to study many aspects of galaxy formation and evolution, including the evolution of specific star formation rates (e.g. \citealt{ilbert15}), effects of environment on galaxy morphology (e.g. \citealt{capak07b}), high-redshift quasars (e.g. \citealt{masters12}) and dust obscured galaxies (e.g. \citealt{riguccini15}). However, the multiwavelength dataset was processed with different flux measurements and reduction methods resulting in a corresponding increase in systematic error.

Here, we construct a catalogue of consistent total flux measurements spanning from the far-ultraviolet to the far-infrared for a subregion we shall refer to as G10 and based on existing COSMOS imagery. Our catalogue, when combined with the spectroscopic redshifts curated by \citet{davies15}, forms an intermediate redshift sample prepared in an identical way to and thus suitable for direct comparison to GAMA. The combined multiwavelength dataset is able to sample multiple processes occurring in the galaxy population across $0 < z < 1$, including (rest frame) ultraviolet light from star formation, optical and near-infrared emission from young and old stars, mid infrared emission from polycyclic aromatic hydrocarbons and warm dust (50~K), and far-infrared emission from cold dust (20~K). In Section \ref{sec:data}, we describe the multiwavelength dataset used. In Section \ref{sec:phot}, we construct a consistent 38 band photometric catalogue spanning the far-ultraviolet to the far-infrared in a subset of the COSMOS region using \textsc{lambdar}. In Section \ref{sec:con}, we demonstrate consistency with existing photometric catalogues in the region. Sections \ref{sec:release} and \ref{sec:conclusion} summarise the release content and present concluding remarks respectively. In four upcoming papers we use this data in conjunction with GAMA to examine stellar and dust masses (Driver et al. 2016 in prep, Wright et al. 2016 in prep.), the cosmic spectral energy distribution (Andrews et al. 2016 in prep) and star formation rates (Davies et al. 2016 in prep).

We use AB magnitudes throughout this work. 

\section{Data}
\label{sec:data}

\begin{figure*}
\begin{minipage}{7in}
\begin{center}
\includegraphics[width=0.99\linewidth]{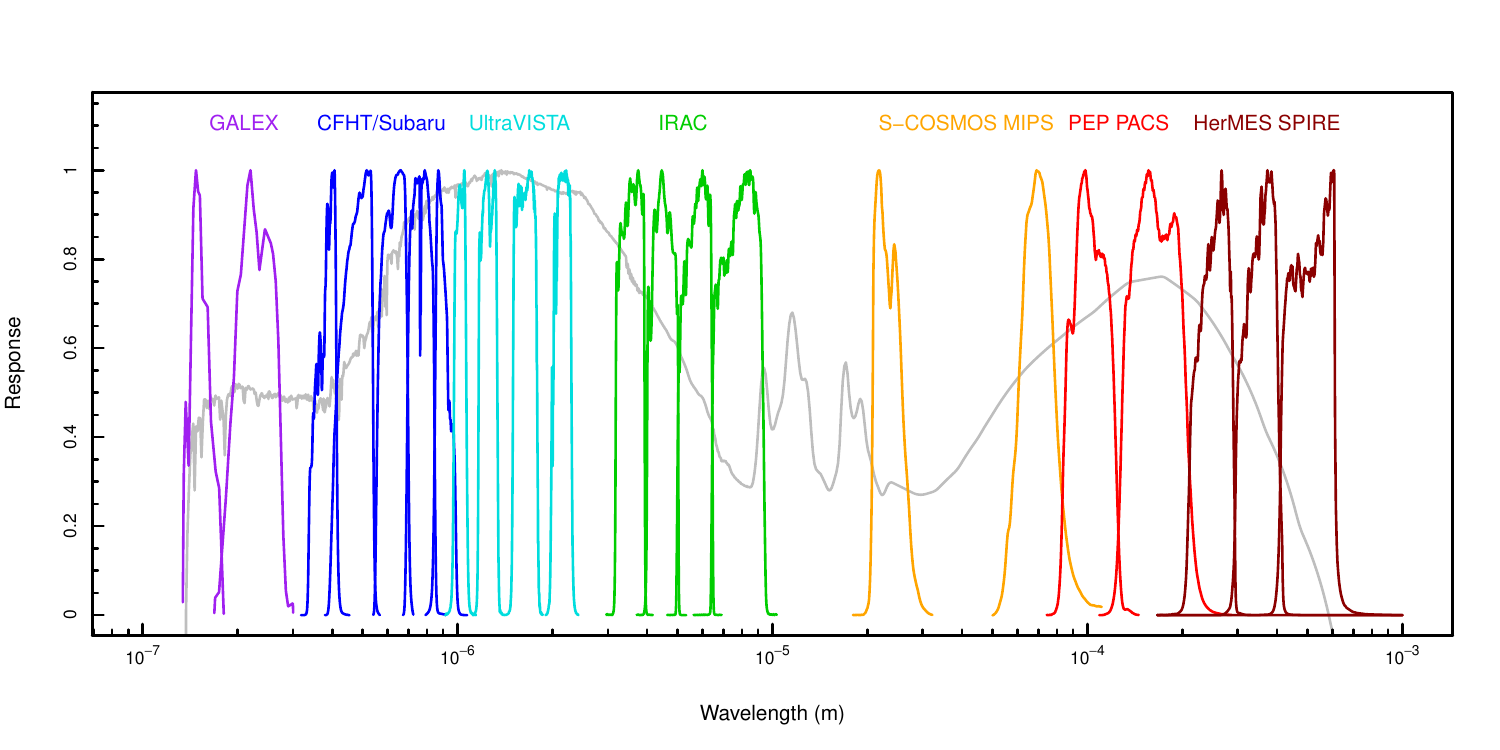}
\caption{The combined broadband filter curves of this multiwavelength dataset, colour-coded by survey and normalised to 1. Subaru $B$, $V$ and the intermediate and narrow bands are omitted for clarity. In grey is the \citet{driver12} cosmic spectral energy distribution redshifted to $z=0.5$ to illustrate what the energy-weighted average galaxy spectral energy distribution may look like at this redshift.}
\label{fig:filterset}

\includegraphics[width=0.99\linewidth]{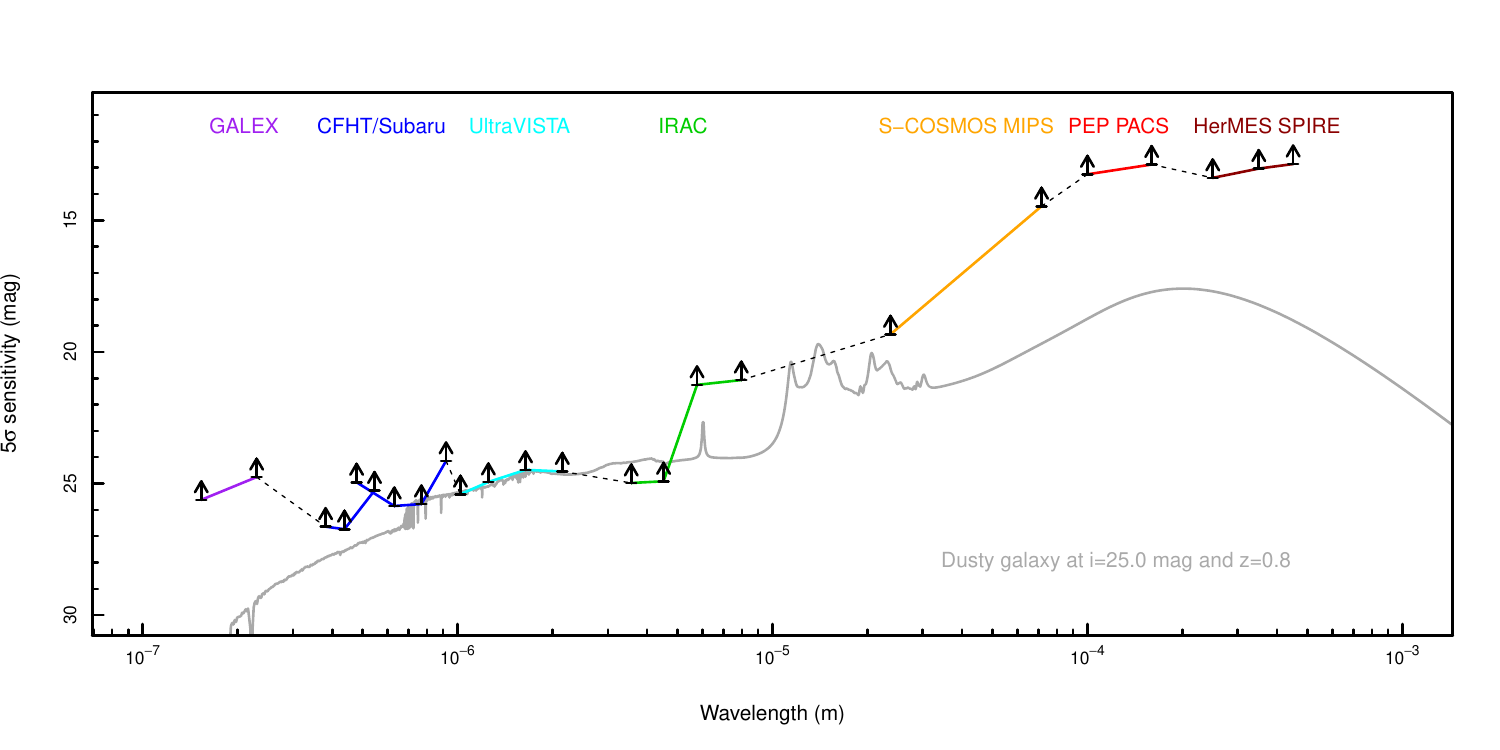}
\caption{The 5$\sigma$ limit for point sources of the different multiwavelength components as computed from the median sky RMS for 1000 random sources in the respective bands for the broadband filters. The grey curve is an example SED fit of a dusty galaxy, scaled to $i = 25~$mag and redshifted to $z = 0.8$. }
\label{fig:sky}

\end{center}
\end{minipage}
\end{figure*}

In this section, we describe the imaging and redshift information used to construct our multiwavelength catalogue and their respective surveys. Figure \ref{fig:filterset} shows the combination of 22 broadband filters used in these surveys (omitting $B$, $V$ and the intermediate and narrow bands for clarity). Also shown, for illustrative purposes only, is the \citet{driver12} cosmic spectral energy distribution redshifted to $z=0.5$ to highlight the emission from a typical galaxy (barring evolution, which will be examined in a later study). The 5$\sigma$ point source detection thresholds of these datasets, as computed from the median sky RMS for 1000 random sources, are shown in Figure \ref{fig:sky}. 

\subsection{GAMA G10}

\begin{figure}
\begin{center}
\includegraphics[width=0.99\linewidth]{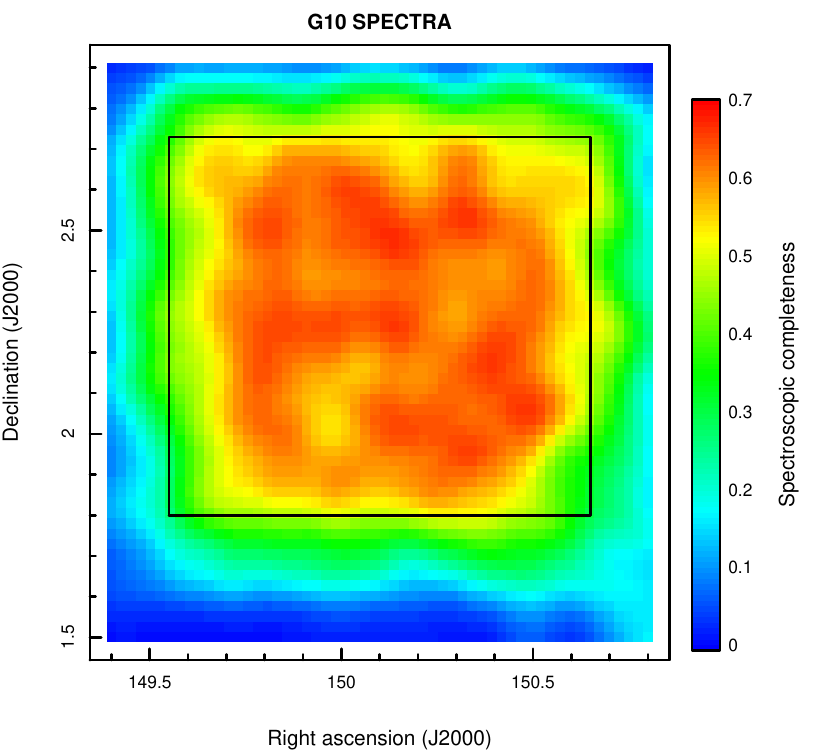}
\caption{Spectroscopic completeness of the COSMOS field for $i < 22.0$~mag on 1.5\arcmin scales. The G10 region is denoted by the black box.}
\label{fig:complete}
\end{center}
\end{figure}

To provide an intermediate redshift comparison for the GAMA project, \citet{davies15} re-reduced spectra from zCOSMOS \citep{lilly07,lilly09} and combined them with spectroscopic redshifts from other surveys, obtaining redshifts for over 22000 sources. The ``bright" component of zCOSMOS targeted 20,000 sources closer than $z < 1.2$ using the VISible Multi-Object Spectrograph (VIMOS) on the Very Large Telescope (VLT). The zCOSMOS observations used a slit length of 10\arcsec, which raises the possibility of confusion in dense regions.

The zCOSMOS-bright raw spectra were re-reduced using a bespoke pipeline, fitted using \textsc{autoz} \citep{baldry14} and position matched to a reverse engineered version of the non-public zCOSMOS input catalogue. The mismatch rate between the original catalogue and the reverse-engineered catalogue is estimated to be 2 per cent.

Both the \textsc{autoz} and original zCOSMOS redshifts were then combined with spectroscopic redshifts from PRIMUS, VVDS and SDSS and photometric redshifts from \citet{ilbert09}, and matched to the broader COSMOS photometric catalogue. As a result of this combination, each source in the COSMOS catalogue is automatically assigned both a ``best" redshift (Z\_BEST) and a reliability flag (Z\_USE) --- Z\_USE = 1 indicates high-resolution, reliable spectroscopic redshifts and Z\_USE $< 3$ represents reliable spectroscopic redshifts. Spectra for all zCOSMOS targeted sources were then visually inspected with the redshift and reliability flag adjusted accordingly.

The G10 region (R.A. $\in [149.55\degr, 150.65\degr$], DEC $\in [1.80, 2.73\degr$]) is a subset of the COSMOS region chosen for its relatively high spectroscopic completeness of $\sim45$ per cent for extra-galactic sources with $i < 22$~mag. This spectroscopic completeness is shown in Figure \ref{fig:complete}. The G10 region has full multiwavelength coverage except for the UltraVISTA bands (0.25 per cent missing). This work uses the second version of the \citet{davies15} catalogue, G10CosmosCatv02, as described in Section \ref{sec:release}.\footnote{\url{http://cutout.icrar.org/G10/dataRelease.php}}

\subsection{COSMOS}
As part of the COSMOS multi-wavelength imaging campaign, \citet{capak07,taniguchi07} and \citet{taniguchi16} obtained imagery from the 8.3~m Subaru telecope in $BgVriz$ and 14 narrow and intermediate bands and the 3.6~m CFHT in the $ui$ bands. Both telescopes are situated on Mauna Kea, Hawaii.

The Subaru imaging was obtained using Suprime-Cam \citep{komiyama03} in 2004 and 2005 with exposure times ranged from 5.8~h to 10.8~h. Suprime-Cam is an array of 10 $2048 \times 4096$ CCDs with a $34\arcmin \times 27\arcmin$~field of view and native resolution of 0.202\arcsec~pixel$^{-1}$. Worst case seeing for the Subaru data ranged from 0.95\arcsec~in $B$ and $i$ to 1.58\arcsec~in $g$. \citet{taniguchi07} claims 90 per cent completeness in $BgVriz$ for exponential disk galaxies down to 24.7, 24.3, 24.1, 24.1, 23.5 and 22.9~mag respectively. A followup survey in 2006 and 2007 (presented by \citealt{taniguchi16}) added imagery in 13 intermediate and narrow band filters.

The CFHT images were obtained using the Mega-Prime camera \citep{aune03,boulade03} from 2003 December to 2005 June and combined using \textsc{swarp} \citep{bertin02}. Mega-Prime is an array of 36 2K $\times$ 4.5K EEV CCDs with a 1~deg$^2$ field of view with a native resolution of 0.18\arcsec~pixel$^{-1}$. The worst case seeing was 0.9\arcsec~in $u$ and 0.94\arcsec~in $i$ and the $5\sigma$ limiting magnitude for a 3\arcsec~circular aperture was 26.5~mag in $u$ and 23.5~mag in $i$. 

Images from both telescopes were resampled to a common resolution of 0.15\arcsec~pixel$^{-1}$ and aligned on to a common astrometric grid by the COSMOS collaboration. This work uses the original point spread function (PSF) (Subaru version 2) mosaics for all bands\footnote{\url{https://irsa.ipac.caltech.edu/data/COSMOS/images/subaru/}} except CFHT $u$ which contains a zeropoint error. In this case, we assembled a mosaic covering the entire COSMOS region from the individual original PSF (CFHT version 5) tiles available on the COSMOS archive\footnote{\url{https://irsa.ipac.caltech.edu/data/COSMOS/images/cfht/}} using \textsc{swarp}.

\subsection{GALEX}
The COSMOS region was observed using the \textit{Galaxy Evolution Explorer} (\textit{GALEX}) as part of the Deep Imaging Survey \citep{zamojski07}\footnote{\url{https://irsa.ipac.caltech.edu/data/COSMOS/images/galex/}}. \textit{GALEX} was an ultraviolet space observatory operated by NASA, launched on 2003 April 28 and decommissioned on 2013 June 28. The observatory was equipped with a 0.5~m mirror, a circular field of view 1.2\degr~in diameter, a 1.5\arcsec~pixel$^{-1}$ detector and two passbands in the far and near ultraviolet (FUV and NUV) respectively. Observations consisted of four pointings with exposure times of 45000~s in FUV with a PSF full width at half maximum (FWHM) of 5.4\arcsec~and 50000~s in NUV with a PSF FWHM of 5.6\arcsec. We assembled the four \textit{GALEX} (version 2) pointings for each band into a single mosaic using \textsc{swarp} with background subtraction turned off.

\subsection{UltraVISTA}
The near-infrared UltraVISTA survey \citep{mccracken12} is currently being conducted on the Visible and Infrared Survey Telescope for Astronomy (VISTA) using the VISTA Infrared Camera (VIRCAM). VISTA is a 4~m telescope operated by the European Southern Observatory in Paranal, Chile. VIRCAM is an array of 16 Raytheon 2048 $\times$ 2048 CCDs with a native resolution of 0.339\arcsec~pixel$^{-1}$.

This work uses the second UltraVISTA data release\footnote{\url{http://ultravista.org/DR2}}, which surveyed the COSMOS region during 2009 December to 2012 May in the $YJHK_s$ wideband filterset for at least 11.1, 12.8, 13.3 and 10.6~h. The typical PSF FWHM was 0.9\arcsec~across all four bands. UltraVISTA is composed of two components -- Deep and Ultra-deep surveys -- outlaid on the sky in alternating vertical stripes $\sim0.2\deg$ wide in RA. The Deep survey claims limiting magnitudes of 25.1, 24.8, 24.4 and 24.5~mag in $YJHK_s$, while the Ultra-deep survey (as of DR2) claims limiting magnitudes of 25.7, 25.4, 25.0, 24.8~mag respectively. A small portion of the G10 region near RA=$150.55\degr$, DEC=$1.83\degr$ has no data in any UltraVISTA band due to a faulty detector (see Figure \ref{fig:mogs}). UltraVISTA DR2 images have been resampled to a pixel scale of 0.15\arcsec~pixel$^{-1}$ and aligned to the COSMOS astrometric grid by the UltraVISTA collaboration.

\subsection{S-COSMOS and SPLASH}

The COSMOS \textit{Spitzer} survey (S-COSMOS; \citealt{sanders07}) surveyed the COSMOS region using the \textit{Spitzer} space telescope. \textit{Spitzer} is a 0.85~m mid-infrared space observatory operated by NASA launched on 2003 August 25. S-COSMOS observed in all passbands of the Infrared Array Camera (IRAC) and the Multiband Imaging Photometer for \textit{Spitzer} (MIPS). After exhaustion of the cryogen on 2009 May 15 only the two shortest wavelength IRAC passbands are operational. 

IRAC is a set of two 256 $\times$ 256 pixel detectors with four filters centred on 3.6, 4.5, 5.6, 8.0~$\mu$m referred to as bands 1 through 4 respectively. IRAC has a native pixel resolution of 1.2\arcsec~pixel$^{-1}$ and field of view of 5.2\arcmin~$\times$5.2\arcmin. IRAC G03 observations of the COSMOS region have a typical exposure time of 1200--2200~s and PSF FWHMs of 1.7\arcsec, 1.7\arcsec, 1.9\arcsec~and 2.0\arcsec~in bands 1 through 4 respectively. 

During the \textit{Spitzer} warm mission, the \textit{Spitzer} Large Area Survey with Hyper-Suprime-Cam (SPLASH; Capak et al. 2016, in prep.) surveyed the COSMOS field with a typical exposure time of 3.8~h per pixel in IRAC channels 1 and 2. These observations achieved a $5\sigma$ depth of 0.2~$\mu$Jy, compared to 0.9~$\mu$Jy for S-COSMOS. The released images have been resampled to 0.6\arcsec~pixel$^{-1}$ by the S-COSMOS and SPLASH collaborations. 

MIPS was a set of three detector arrays with 128 $\times$ 128, 32 $\times$ 32 and 2 $\times$ 2 pixels with a pixel scale of 1.2, 4.0 and 8.0\arcsec~pixel$^{-1}$ at 24, 70, and 160 $\mu$m respectively. The G03 MIPS observations of COSMOS took place during 2006 January to 2008 January. Integration times were 2800, 1350 and 270~s and PSF FWHMs were 5.9\arcsec, 18.6\arcsec~and 39\arcsec~for 24.0, 70.0, and 160.0 $\mu$m respectively. The $1\sigma$ noise level was 1.7 and 13 mJy in 70~$\mu$m and 160~$\mu$m. The MIPS 70~$\mu$m and 160~$\mu$m observation strategy and data reduction process is described in \citet{frayer09} and the MIPS 24 catalogue is briefly described in \citet{lefloch09}.

This work uses all SPLASH data and MIPS observations at 24 (version 1) and 70 $\mu$m (version 3)\footnote{\url{https://irsa.ipac.caltech.edu/data/COSMOS/images/spitzer/mips/}}. We do not adopt the MIPS 160 data as \textit{Herschel} data offers superior sensitivity and resolution.

\subsection{PACS Evolutionary Probe}
The PACS (Photodetector Array Camera and Spectrometer) Evolutionary Probe (PEP; \citealt{lutz11}) was a survey conducted on the \textit{Herschel} space observatory. \textit{Herschel} \citep{pilbratt10} was a 3.5~m far-infrared space telescope operated by the European Space Agency from launch on 2009 May 14 to 2013 April 29, when the cryogenic coolant was exhausted. 

PACS \citep{poglitsch10} was a combined imagery and integral field spectroscopy instrument whose two 16 $\times 25$ pixel bolometer arrays had pixel scales of 1.2 and 2.4\arcsec~pixel$^{-1}$ and a field of view of 1.75\arcmin~$\times$ 3.5\arcmin. The instrument featured passbands centred around 70, 100 (1.2\arcsec~pixel$^{-1}$) and 160~$\mu$m (2.4\arcsec~pixel$^{-1}$). 

This work uses the first PEP data release\footnote{\url{http://www.mpe.mpg.de/ir/Research/PEP/DR1}}. PEP surveyed the COSMOS region at 100~$\mu$m and 160~$\mu$m for 196.9~h in the period 2009 November -- 2010 June, yielding images with PSF FWHM of 7.4\arcsec~and 11.3\arcsec~at 100~$\mu$m and 160~$\mu$m respectively. The observations achieved a confusion-limited $3\sigma$ sensitivity at 160~$\mu$m of 10.2~mJy.

\subsection{HerMES}

The \textit{Herschel} Multi-Tiered Extragalactic Survey (HerMES; \citealt{oliver12}) was a far-infrared survey conducted on the \textit{Herschel} space observatory using the Spectral and Photometric Imaging Receiver (SPIRE). SPIRE \citep{griffin10} was a combined three-band imager and Fourier-transform spectrometer with a 4\arcmin $\times$ 8\arcmin~field of view. The imaging bands were centred on approximately 250, 350 and 500~$\mu$m with pixel scale 6.0, 8.3 and 12.0\arcsec~pixel$^{-1}$ and FWHM of 18.15\arcsec, 25.15\arcsec~and 36.3\arcsec~respectively. 

This work uses the second HerMES data release\footnote{\url{http://hedam.lam.fr/HerMES/}}. HerMES surveyed the COSMOS region in the three SPIRE bands for approximately 50~h, achieving a $5\sigma$ noise limit of 8.0, 6.6 and 9.6~mJy in 250, 350 and 500 $\mu$m respectively. The reduction process for the HerMES images is described in \citet{levenson10} and \citet{viero13} and the construction of the HerMES catalogues is described in \citet{roseboom10,smith12} and \citet{wang14}. 

\subsection{Characteristics of the multiwavelength dataset}

\begin{figure*}
\begin{minipage}{7in}
\begin{center}
\includegraphics[width=0.99\linewidth]{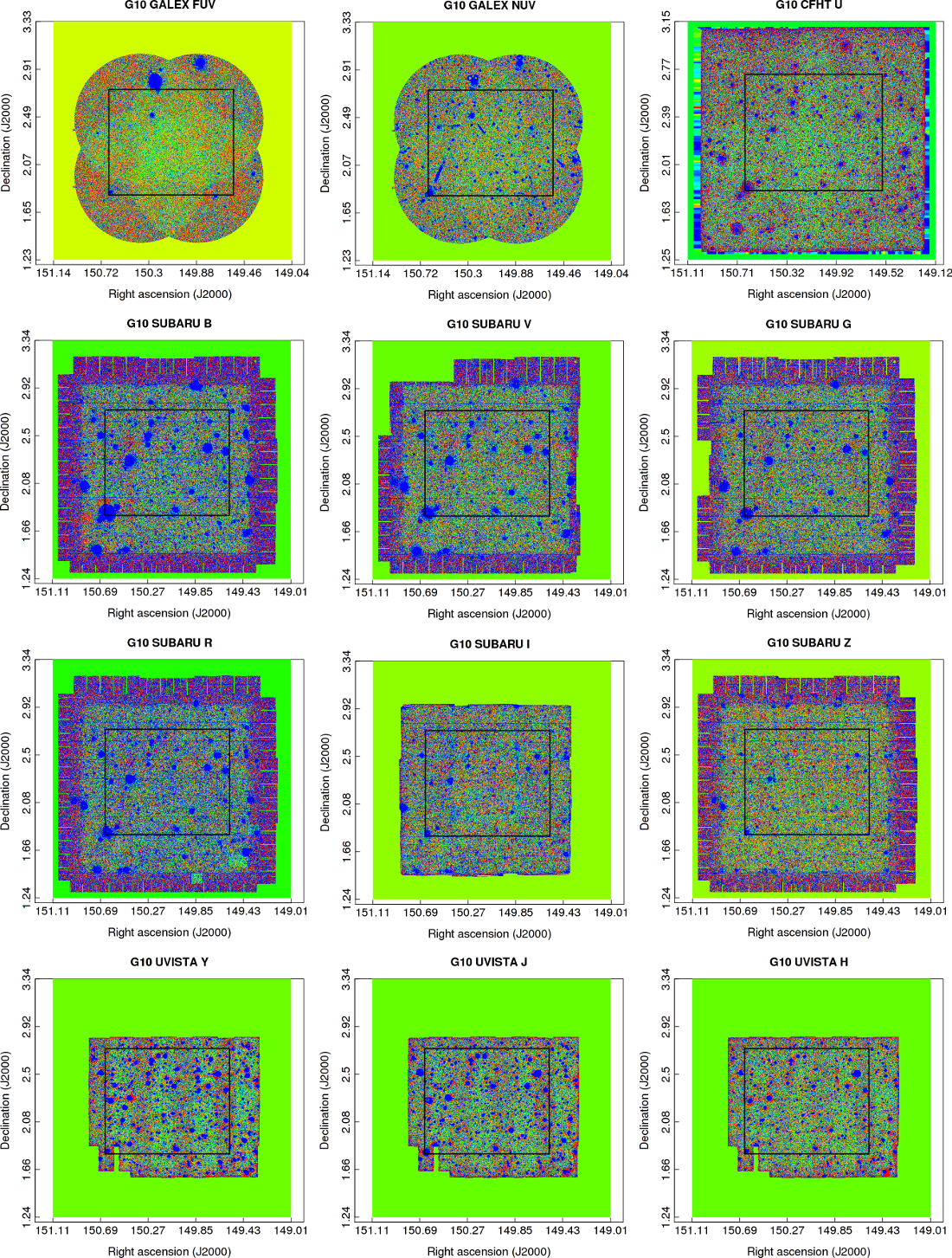}

\caption{Background uniformity and coverage for the \textit{GALEX}, COSMOS, UltraVISTA, S-COSMOS, PEP and HerMES data. The black box denotes the G10 region. Resolution has been reduced to 1.5\arcsec per pixel and levels adjusted to be close to the sky noise. }
\label{fig:mogs}
\end{center}
\end{minipage}
\end{figure*}

\begin{figure*}
\begin{minipage}{7in}
\begin{center}
\includegraphics[width=0.99\linewidth]{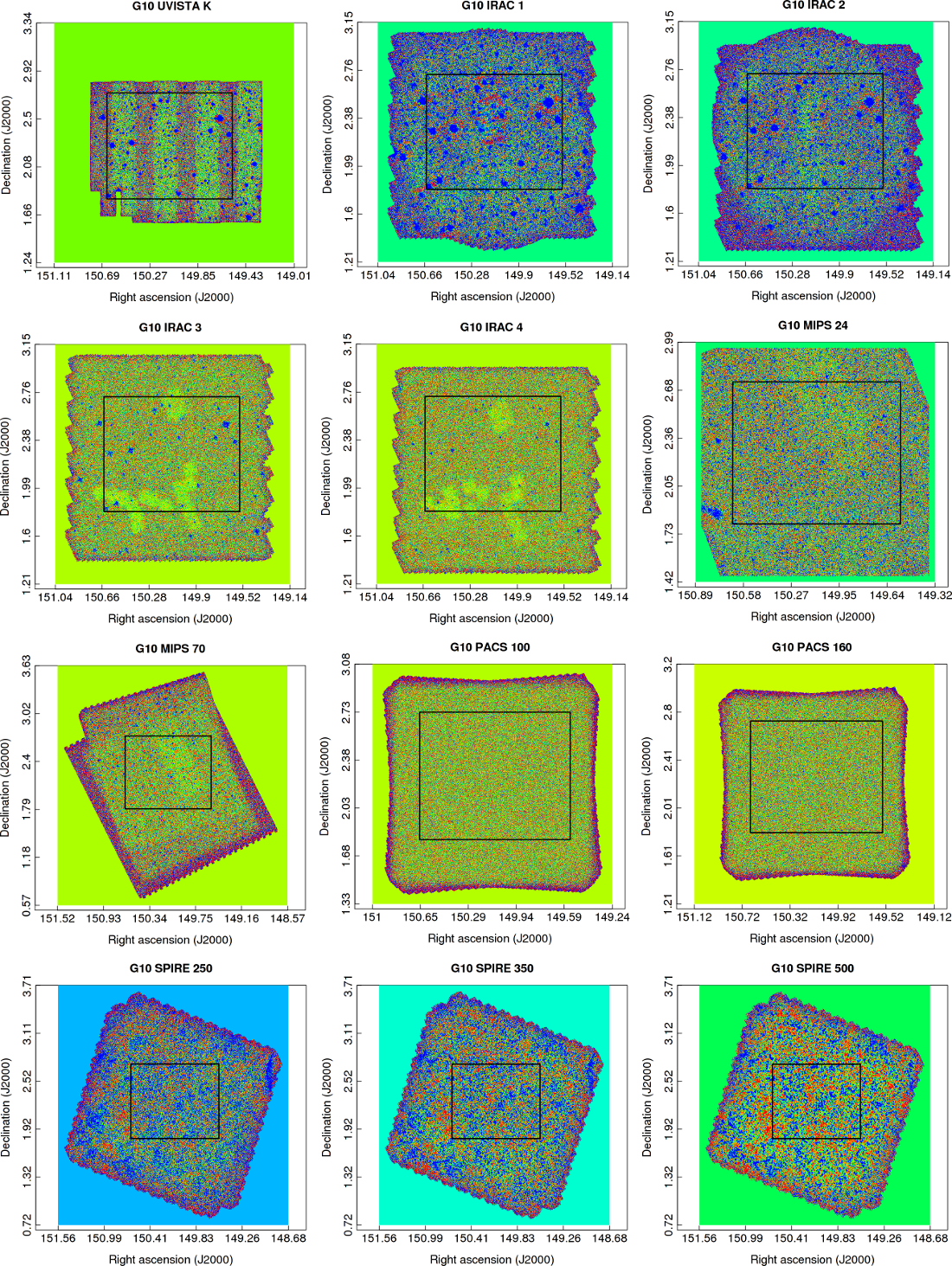}
\contcaption{}
\end{center}
\end{minipage}
\end{figure*}

Figure \ref{fig:mogs} presents an overview of the coverage and background properties of the multiwavelength dataset. Images were resampled to 1.5\arcsec per pixel using \textsc{swarp} and levels adjusted to be close to the sky background ($\pm 2\sigma$) using \textsc{mogrify}. At these scales, astronomical objects are not visible. The circular ``holes" in wavelengths shorter than IRAC 4 are halos of faint light surrounding saturated stars. The four circular pointings of \textit{GALEX} and the alternating deep and ultradeep stripes of UltraVISTA are both clearly visible. Also visible are window and dichroic reflections (rings) and bevel reflections (streaks) in the \textit{GALEX} NUV data. The non-uniform background seen in the HerMES data is a result of Galactic cirrus.

Both the G10 catalogue and a cutout generator for the multiwavelength imagery described above are available at \url{http://cutout.icrar.org/G10/dataRelease.php}. Some key details of the multiwavelength imagery are outlined in Table \ref{tab:settings}.

\section{Photometry}
\label{sec:phot}

\subsection{Existing photometry in the G10 region}

Existing photometry in the COSMOS region consists of an assortment of independent flux measurements and data reduction methods. For instance, the \citet{capak07} optical COSMOS photometric catalogue provides an AUTO flux measurement in only the Subaru $i$ band and fixed 3\arcsec apertures for all Subaru and CFHT bands, UltraVISTA provides SExtractor \citep{bertin96} derived AUTO magnitudes for all four near-infrared bands and S-COSMOS provides a catalogue of four different flux measurements using fixed sized apertures (1.4\arcsec, 1.9\arcsec, 2.9\arcsec and 4.1\arcsec) in the IRAC bands. \citet{muzzin14} homogenizes this dataset, but they calculate 2.1\arcsec fixed size apertures and do not include the \textit{Herschel} data. 

Inhomogenous analytical techniques and the use of fixed size apertures can potentially introduce systematic and random errors for a subset of scientific investigations such as measuring the cosmic spectral energy distribution at intermediate redshifts. Figure \ref{fig:shredded} shows that the 2 and 3\arcsec apertures used by the COSMOS 2007 (now obsolete) catalogue are unsuitable for objects at low to intermediate redshifts. 3\arcsec apertures are also unsuitably large for the region depicted in the right panel. 

Recently, the COSMOS2015 catalogue \citep{laigle16} derived fixed-size and AUTO aperture photometric measurements for sources originally detected on a $zYJHK$ co-added image. Aperture definitions were propagated to PSF matched images from $u$ through $K$ using \textsc{SExtractor}'s dual image mode. The catalogue also contains existing photometry from \textit{GALEX}, updated photometry from IRAC derived using IRACLEAN \citep{hsieh12}, updated MIPS photometry and \textit{Herschel} photometry based on MIPS 24~$\mu$m priors, but does not include MIPS 70 or $g$. 

While the COSMOS2015 catalogue provides exquisite PSF matched photometry from $u-K$ the full panchromatic grasp is obtained from table matching with previously constructed \textit{GALEX} and \textit{Herschel} catalogues. For the purpose of full SED modelling table matching, as opposed to ``forced'' photometry can introduce unphysical discontinuities at the wavelength interfaces. In SED modelling a consistent measurement and even more importantly a consistent error assessment across the full wavelength range is critical. Whether such issues are relevant can only be established after conducting a fully panchromatic analysis and comparing the outcome. To that extent, the aim of this paper is to re-define a set of detections and apertures and derive consistent, total flux photometry in the G10 region using \textsc{lambdar} specifically for panchromatic analysis over the entire wavelength range. This also permits direct comparisons with the \textsc{lambdar} derived photometry of low-redshift sources from GAMA \citep{wright16a}.

\subsection{LAMBDAR}
We use \textsc{lambdar} \citep{wright16a}\footnote{\url{https://github.com/AngusWright/LAMBDAR}} to construct reliable panchromatic photometry for the G10 region across all bands outlined in Section \ref{sec:data}. \textsc{lambdar} is explicitly designed to deal with the resolution mismatch that arises from multiwavelength datasets, and deblending that is capable of dealing with low resolution and confused far-infrared data. 

Briefly, \textsc{lambdar} requires a set of aperture definitions (RA, DEC, semi-major and semi-minor axes and position angle) and a set of input images. Input images do not have to be pixel-matched nor astrometrically aligned. The input apertures must be robust --- the shredding depicted in the left and centre panels of Figure \ref{fig:shredded} (regardless of the use of fixed sized apertures) will lead to the flux being significantly underestimated, while the incorrect apertures in the right panel of Figure \ref{fig:shredded} would cause flux to be erroneously large for the faint central objects and erroneously small for the surrounding bright objects and objects on the edge of the complex. The user can also supply a list of contaminants --- objects with defined apertures which are fully deblended using the following method, but without flux measurements being performed.

For each image, \textsc{lambdar} optionally convolves input apertures with the PSF, giving an aperture function $A_i (x, y)$ for each object. The PSF may be integrated outward to encapsulate some fraction of the total integral (parameter name \texttt{PSFConfidence}) and truncated at the corresponding radius. The normalization of the aperture functions can be scaled by a set of prior flux weights $w_i$ either supplied by the user or determined using the flux of the central pixel. \textsc{lambdar} then calculates the deblend function

\begin{equation}
D_i (x, y) = \frac{w_i A_i(x, y)}{\sum_i w_i A(x, y)}
\end{equation}

for each object, which is the ratio of the weighted aperture function of the object to the sum of all weighted aperture functions for a particular pixel. The deblended image, i.e. the product of $D_i (x, y)$ and the image $I(x, y)$, governs how much flux is assigned to object $i$. \textsc{lambdar} then performs either PSF weighted photometry --- which multiplies the deblended image by the aperture function --- or aperture photometry, which converts the aperture function into a simple aperture before multiplication. The latter is achieved by integrating $A_i (x, y)$ outward to some fraction of the total integral (\texttt{ApertureConfLimit}) and assigning $A(x, y) = 1$ or 0 accordingly. The process of flux determination may be repeated iteratively, with the output flux measurements being used as input weights. The program is also capable of performing local sky subtraction, blanks and randoms corrections. For full details of the \textsc{lambdar} code, see \citet{wright16a}.

\subsection{Input catalogue}

\newpage

\begin{figure*}
\begin{minipage}{7in}
\begin{center}
\includegraphics[width=0.75\linewidth]{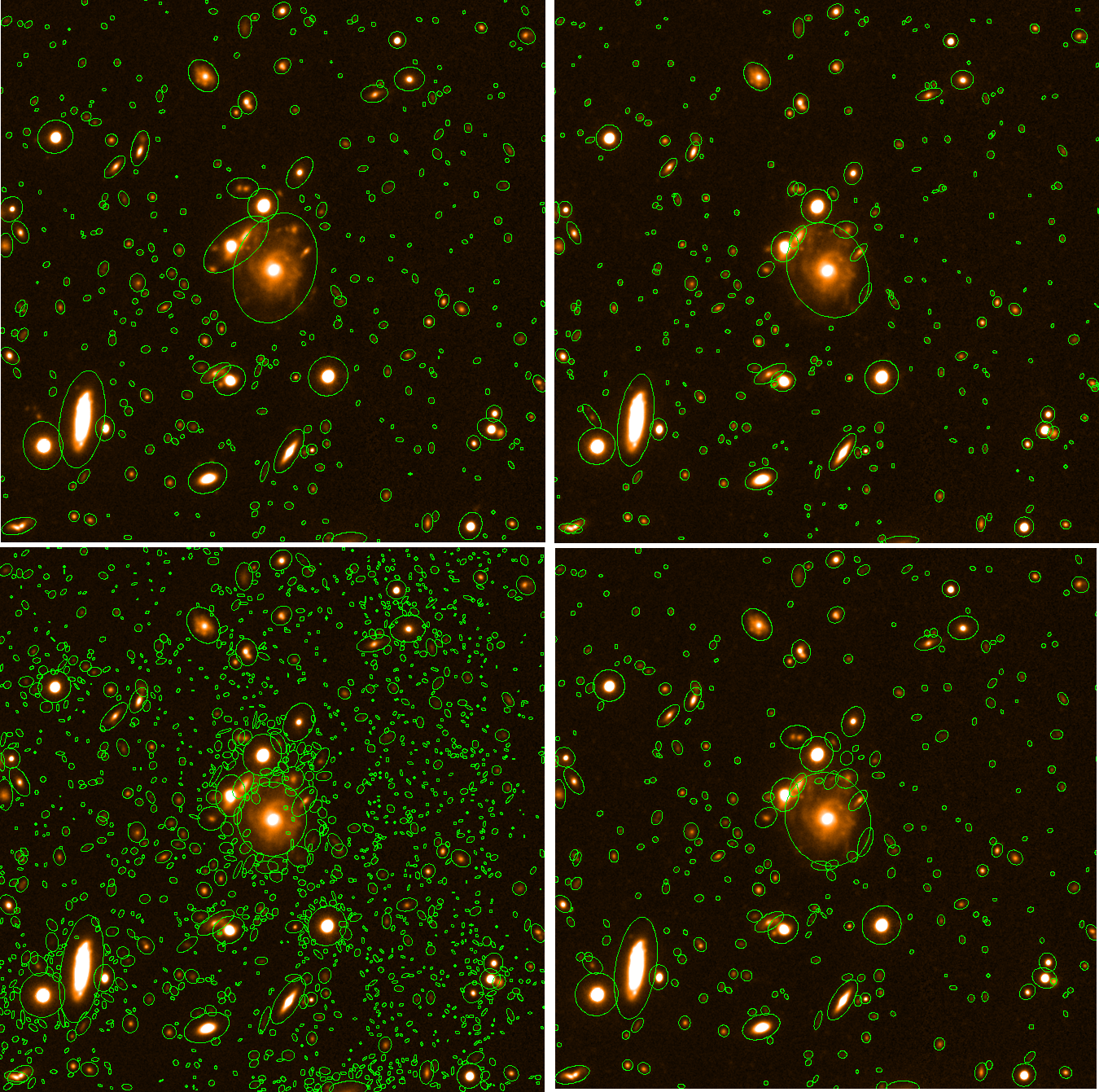}
\caption{A sample region (50'' $\times$ 50'') with apertures obtained using the \textsc{SExtractor} default settings (top left), the COSMOS settings optimised for their PSF matched data (bottom left), the COSMOS2015 settings (bottom right) and our settings (top right). Particular improvement is seen in the deblending solution around the central object, achieving a result similar to COSMOS2015.}
\label{fig:aper}
\end{center}
\end{minipage}
\end{figure*}

\begin{figure*}
\begin{minipage}{7in}
\begin{center}
\includegraphics[width=0.3\linewidth]{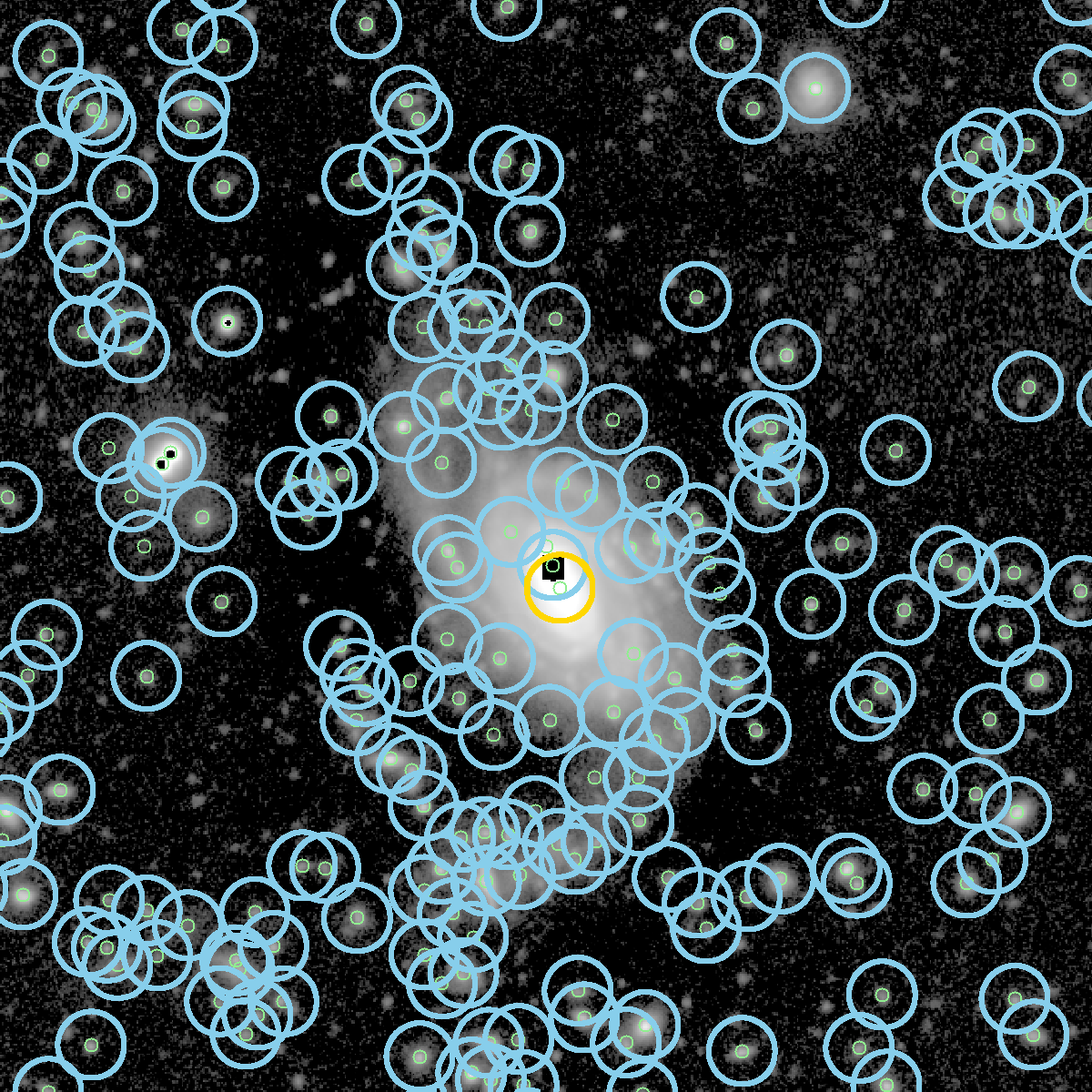}
\includegraphics[width=0.3\linewidth]{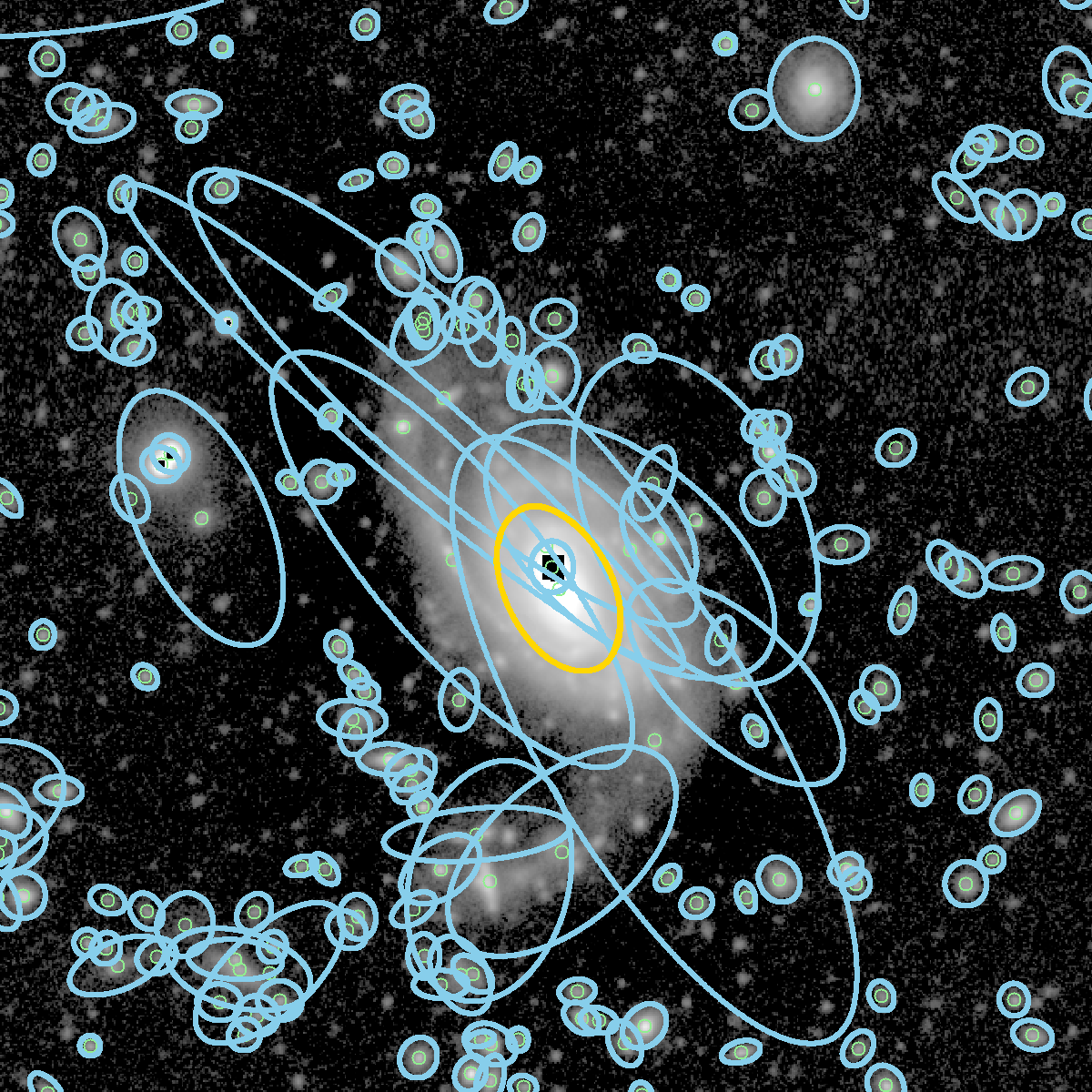}
\includegraphics[width=0.3\linewidth]{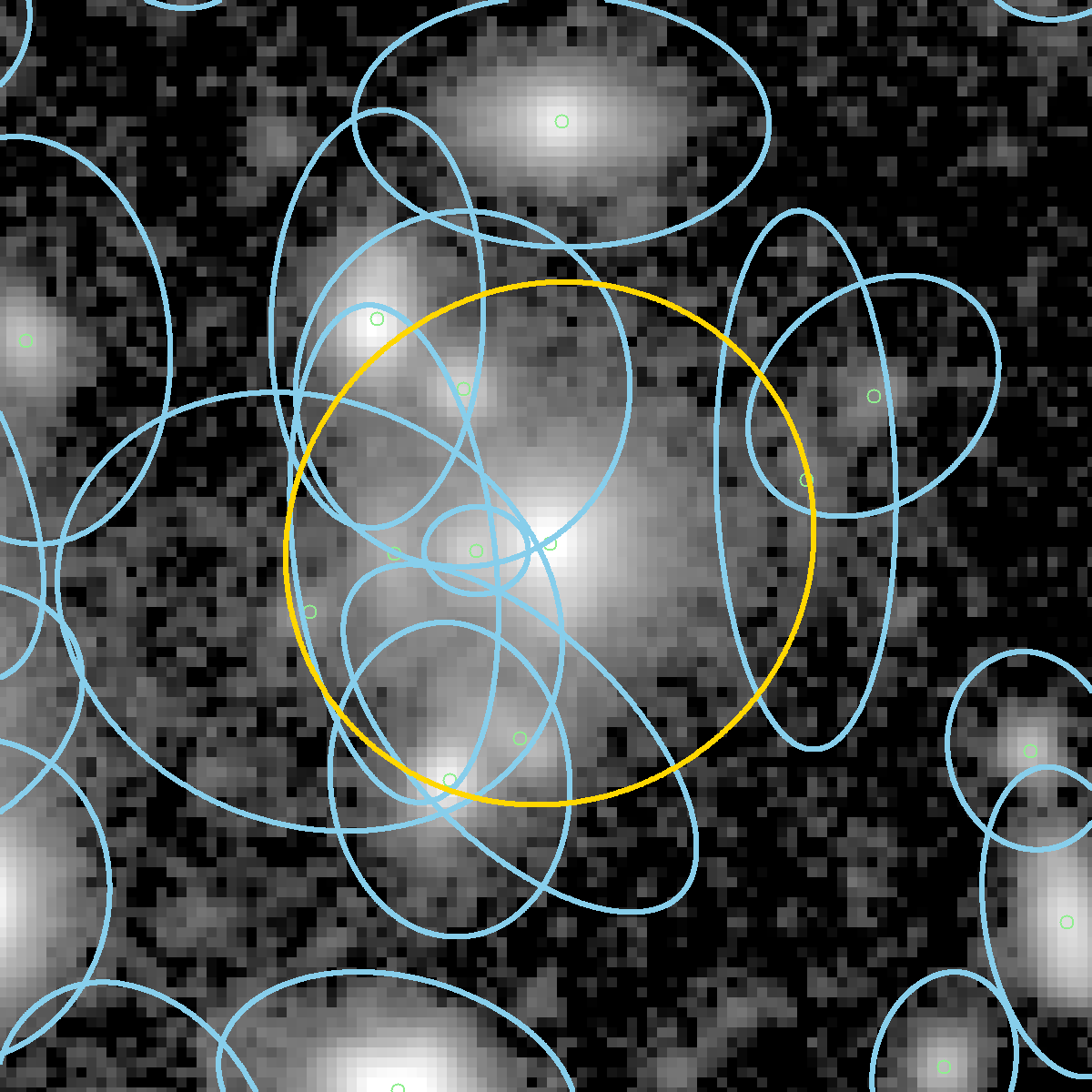}
\caption{Subaru $i$ band cutouts of CATAIDs 6008198 (left and centre, 50'' radius) and 6002104 (right, 8'' radius), denoted by the gold apertures. Left panel: each source is associated with a 3\arcsec aperture and corresponds to an object in the G10/COSMOS photometric catalogue. The centre and right panels show 6008198 and 6002104 respectively, but with aperture parameters derived from our catalogue prior to manual intervention. The saturated region corresponds to a foreground star.}
\label{fig:shredded}
\end{center}
\end{minipage}
\end{figure*}

\newpage

\begin{figure}
\begin{center}
\includegraphics[width=1.0\linewidth]{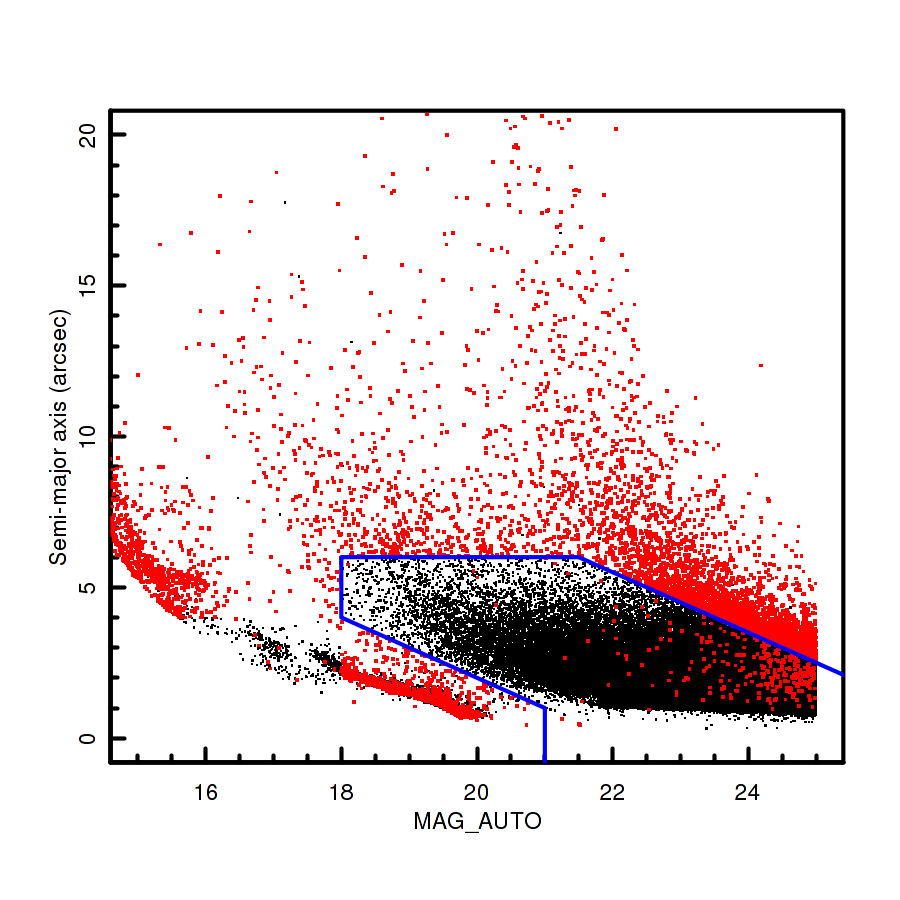}
\caption{\textsc{SExtractor}-derived semi-major axis versus $i$ band magnitude for our sample. Apertures selected for inspection (red) either lie outside the blue lines (see text), or have a positional offset of more than 0.5\arcsec compared to the COSMOS photometric catalogue. Objects outside the blue lines that are not selected for inspection are known spectroscopic stars.}
\label{fig:magsize}
\end{center}
\end{figure}

\begin{figure}
\begin{center}
\includegraphics[width=0.49\linewidth]{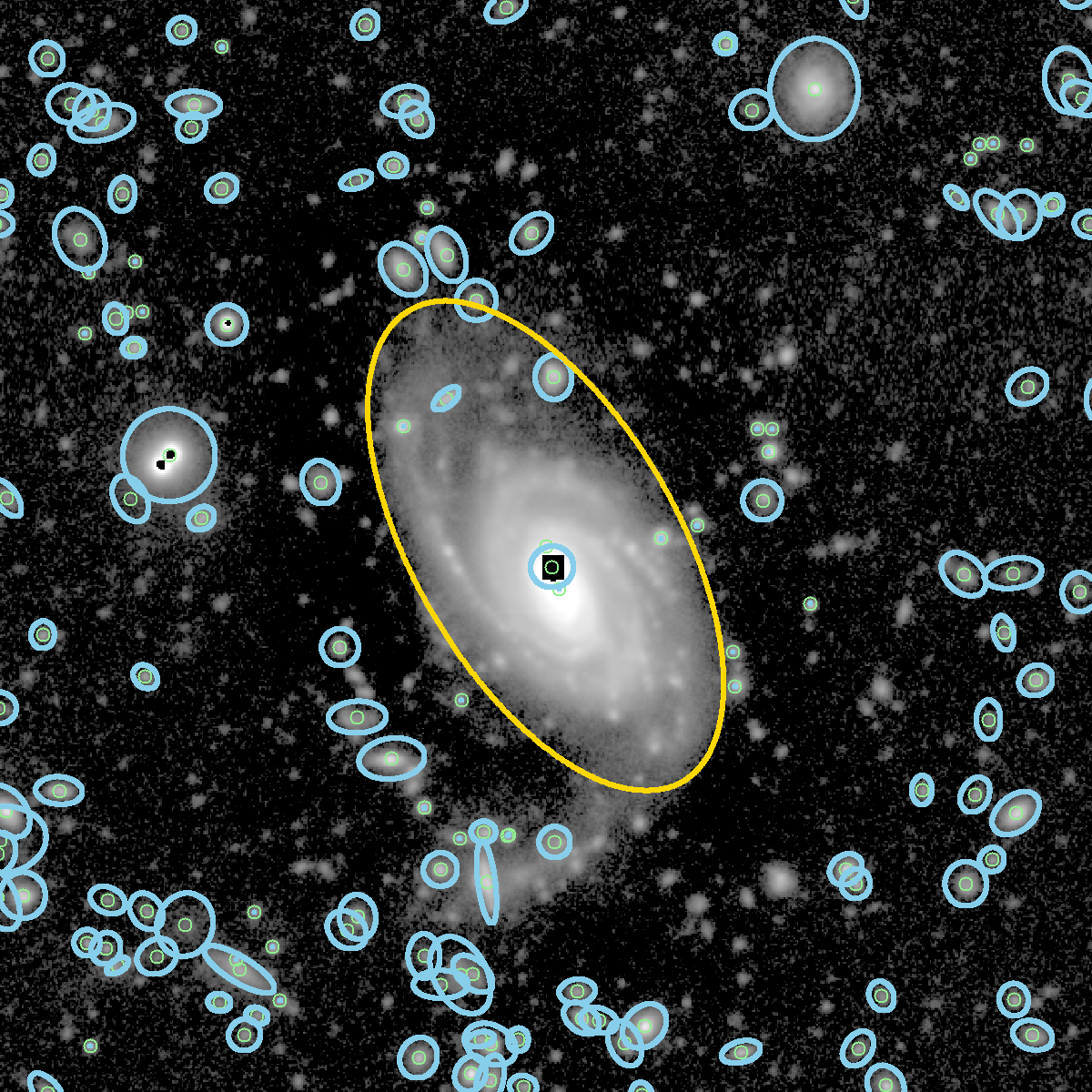}
\includegraphics[width=0.49\linewidth]{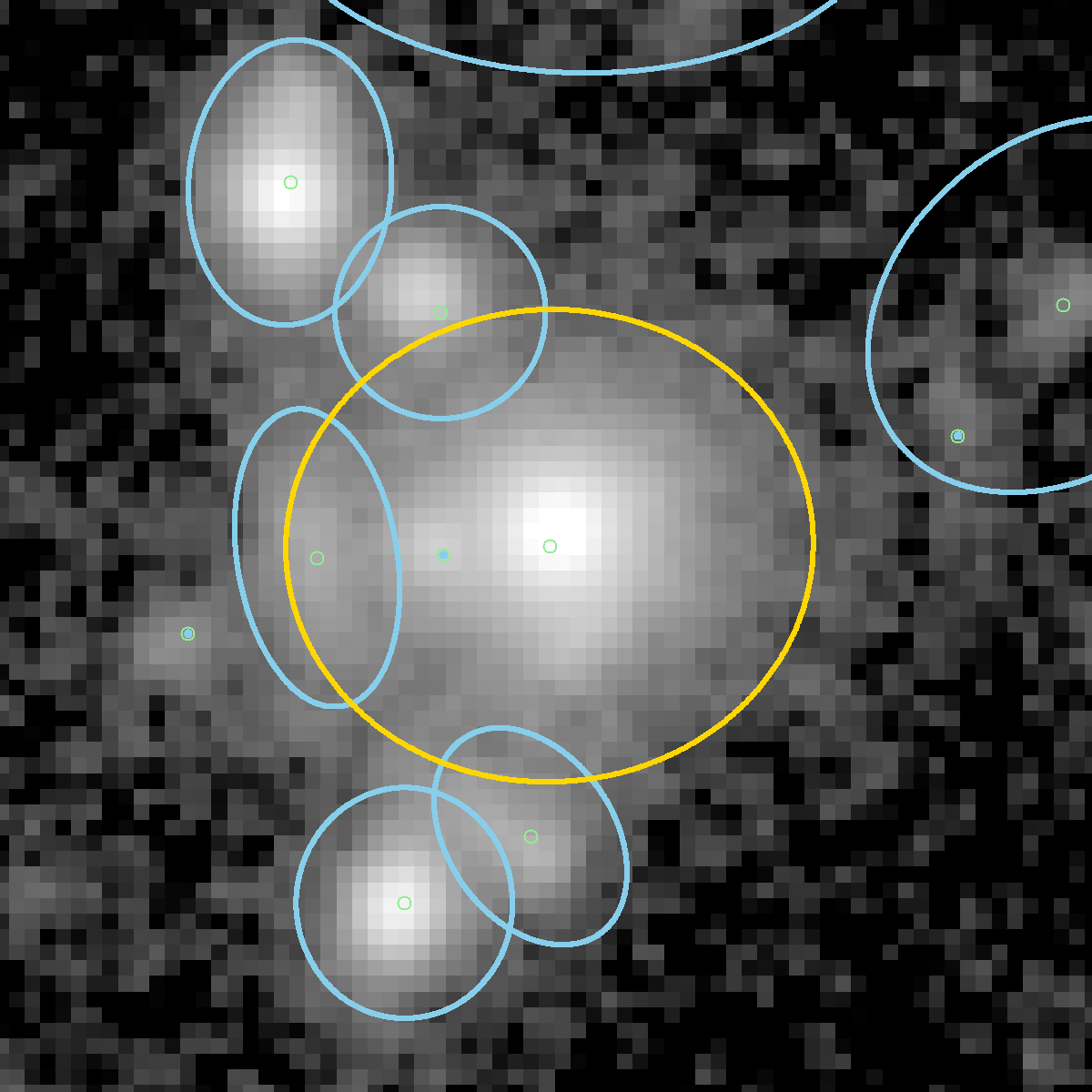}
\caption{Subaru $i$ band cutout of GAMA objects 6008198 (left) and 6002104 (right) using the final aperture catalogue analogous to Figure \ref{fig:shredded}. A green dot without a corresponding aperture ellipse represents a point source.}
\label{fig:unshredded}
\end{center}
\end{figure}

\begin{figure}
\begin{center}
\includegraphics[width=1.0\linewidth]{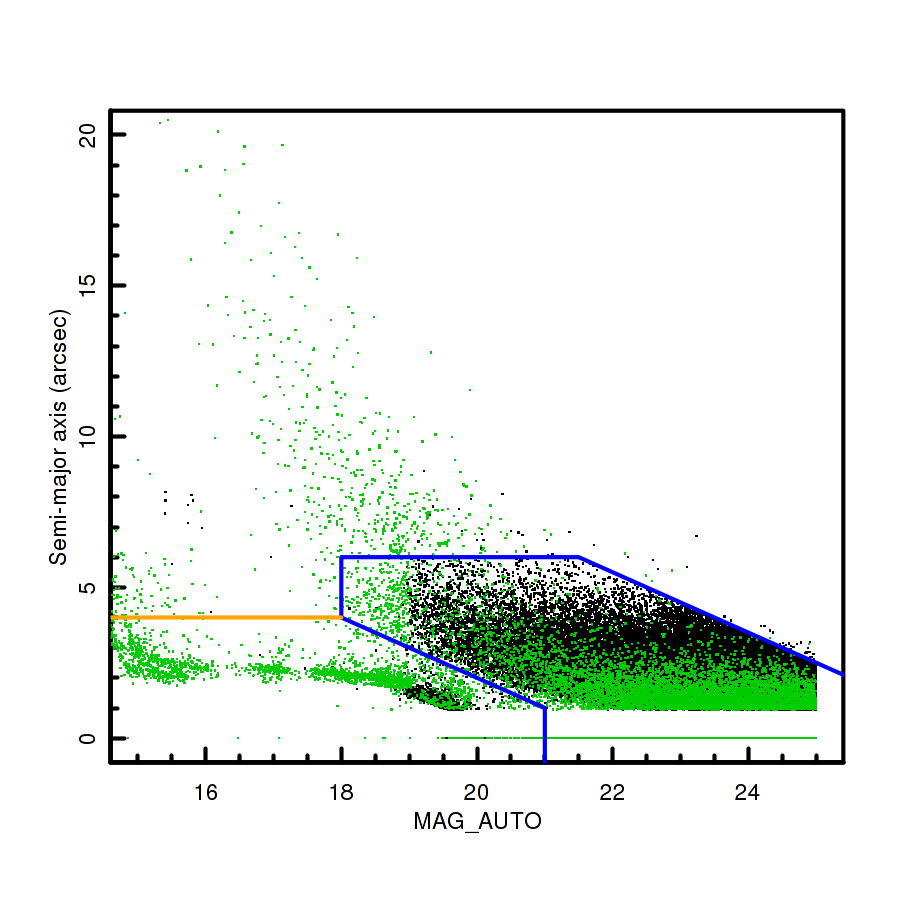}
\caption{Semi-major axis after aperture fixing versus $i$ band magnitude for our sample. Objects with changed apertures are denoted with green points.}
\label{fig:magsize-after}
\end{center}
\end{figure}

As noted above, \textsc{lambdar} does not perform blind source detection and aperture definition. To obtain the input apertures we ran SExtractor (v2.19.5) on the $i$ band Subaru mosaic, with the saturated NaN regions replaced with a nominal value (3001) to avoid shredding of bright stars. After some trial and improvement, we use a detection threshold of 3$\sigma$, analysis threshold of 1.5$\sigma$, deblending parameters DEBLEND\_NTHRESH = 64, DEBLEND\_MINCONT = 0.0004 and no convolution filter. This provides the most qualitatively robust apertures, compromising between faint source detection, close source deblending and minimising over deblending of large resolved sources.  We find the SExtractor default settings do not deblend sufficiently, while the \citet{capak07} settings --- which are tailored for the PSF matched and stacked (CFHT and Subaru $i$ band) image used for the construction of the COSMOS (2007, now obsolete) photometric catalogue --- produce large amounts of false detections and shredded objects (a cut of $i < 25$~mag in total flux was applied to their public catalogue). The COSMOS2015 deblend solution produces similar results to ours. Figure \ref{fig:aper} compares our settings to both the default and COSMOS 2007 (bottom left) and 2015 (bottom right) settings for a small cutout. 

Despite the above choice of parameters, Figure \ref{fig:shredded} (middle panel) shows that problematic apertures still exist for large or flocculent galaxies and near bright objects. We use the aperture magnitude-size plane depicted in Figure \ref{fig:magsize} to identify potentially bad apertures and objects prone to shredding. We then examined cutouts containing objects $i < 25$~mag that meet any of the following criteria: $i < 18$~mag, semi-major axis $> 6\arcsec$, $i$ + semi-major axis (in arcsec, see the lower diagonal line in Figure \ref{fig:magsize}) $< 22$, $i$ + semi-major axis (in arcsec, see the upper diagonal line in Figure \ref{fig:magsize}) $> 27.5$ or a positional offset between the input catalogue and COSMOS photometric catalogue of $> 0.5\arcsec$. These cuts are shown in blue in Figure \ref{fig:magsize}. Apertures above and to the left of this region are either objects that did not deblend correctly or large, bright objects prone to shredding, while non-stellar sources below the region are the result of shredding. For objects that have redshift information, we also required Z\_BEST $> 0.01$ and Z\_USE $< 5$ (sources not flagged as stars). For objects that do not have a redshift measurement, we required objects brighter than $i < 18$~mag to have a semi-major axis $> 4$\arcsec.

Of the 6547 (3.5 per cent of total) sources inspected, 2785 (1.5 per cent) required manual intervention, 651 (0.3 per cent) were false detections and 1838 (1.0 per cent) were replaced with point sources. Manual inspections were performed by SKA, LJMD, and SPD and involved both fixing the primary aperture and fixing, adding or removing apertures down to $i < 25$~mag whose centre lies within 1.5 times the revised semi-major axis of the primary object. Inspections and fixes were performed using a bespoke interface written by ASGR using the \textsc{shiny} framework. In addition, any apertures with a semi-major axis less than the Subaru $i$ PSF FWHM of 0.95\arcsec were replaced with a point source. 

However, the above procedure failed to recover the very brightest stars due to positional mismatch. To overcome this issue, we ran SExtractor on the CFHT $i$ band mosaic with default-like settings --- varying only the detection threshold (3$\sigma$), sky background mesh size (512 pixels) and memory settings. We then position matched this catalogue to the G10/COSMOS catalogue to 1\arcsec and used the CFHT aperture parameters for CATAIDs brighter than $i < 19$~mag (as measured on the CFHT image) that were not manually inspected. This resulted in 2256 apertures (1.2 per cent) being updated.

Another potential issue is that apertures in regions requiring complex deblending were systematically larger than their constituent objects. One such example is presented in the right panel of Figure \ref{fig:shredded}. To (partially) address this problem, we selected objects where the primary aperture has at least five overlapping apertures, or where the weighted sum of the overlapping areas are at least 125 per cent of the area of the primary aperture (an area that is covered by $n$ overlapping apertures is counted $n$ times). We performed an internal match of this list to a 6\arcsec radius, then checked and fixed by eye objects within 6\arcsec or 1.5 times the semi-major axis of the primary object (whichever was larger). 1209 regions were fixed in this manner. The aperture catalogue and mosaic were then overlain on screen and visually inspected, with any obvious remaining problems fixed by hand. 

In total, there are 185907 objects in the G10 region, of which 17062 (9.2 per cent) had apertures requiring manual inspection. This included manually fixing 2785 (1.5 per cent) sources, removing 651 (0.3 per cent) false detections, replacing 1838 (1.0 per cent) with point sources, adding 1722 objects (1.0 per cent) and fixing 1209 (0.7 per cent) regions manually. In this process, 9480 (5.1 per cent) neighbouring sources were also fixed. The resulting size-magnitude distribution, analogous to Figure \ref{fig:magsize}, is shown in Figure \ref{fig:magsize-after} with the changed apertures denoted with green points. Figure \ref{fig:unshredded} is analogous to Figure \ref{fig:shredded}, but with the manual fixes incorporated. Manual inspections took about one minute each, with total time expended on the order of 100 person-hours.

The total number of objects fixed manually, and the number of potentially problematic apertures remaining, highlights the need for increasing the accuracy of automated aperture determination for the next generation of galaxy surveys, for example the Wide Area VISTA Extragalactic Survey (WAVES; \citealt{waves}). These surveys have comparable source density to the COSMOS region but instead cover hundreds of square degrees, making manual intervention prohibitively labour intensive.

\subsection{Obtaining photometry}
\label{sec:phot2}

\begin{figure*}
\begin{minipage}{7in}
\begin{center}
\includegraphics[width=0.99\linewidth]{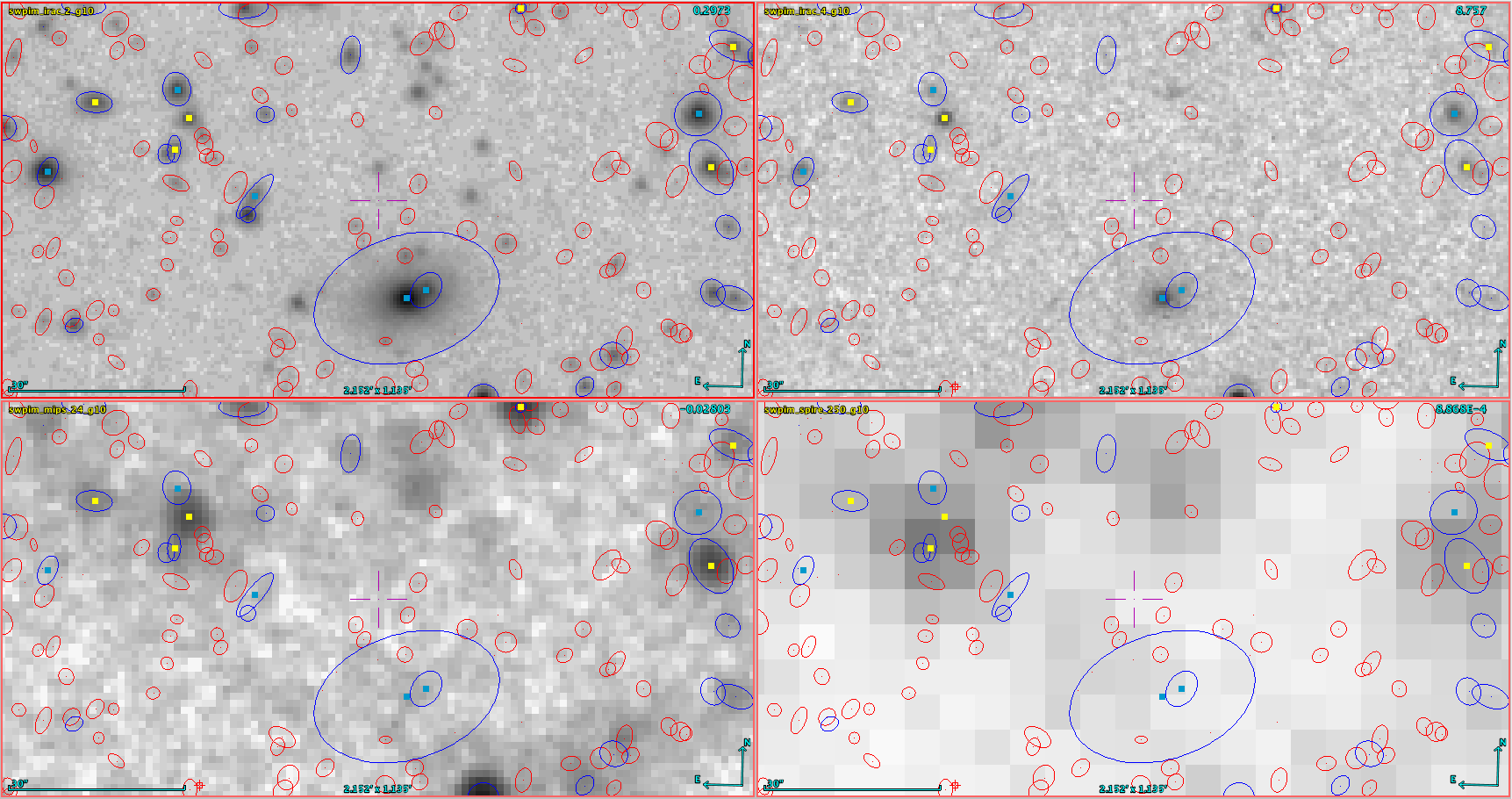}
\caption{Illustration of the cascading flux cats in the mid and far-infrared. Images --- top left: IRAC 2, top right: IRAC 4, bottom left: MIPS 24, bottom right: SPIRE 250. Sources with yellow dots have both MIPS 24 and SPIRE 250 point source photometry and IRAC 4 photometry measured, blue dots have IRAC 4 and MIPS 24 measurements, blue ellipses have IRAC 4 photometry only and red ellipses have none of these. Each cutout spans an area of 2.15' $\times$ 1.14'.}
\label{fig:fluxcuts}
\end{center}
\end{minipage}
\end{figure*}

\begin{table*}
\begin{minipage}{7in}
\begin{center}
\caption{Band-dependant \textsc{lambdar} settings and image metadata}
\begin{tabular}{l|c|c|c|c|c|c|c}
\hline
Band  & PSF FWHM  & Zeropoint & Saturation & Pixel size & Saturation\\
& (\arcsec) & (mag) & (counts) & (\arcsec) & value (counts)\\
\hline
FUV    & See \S\ref{sec:phot2} & 18.82 &   --- &  1.5 & --- \\
NUV    & See \S\ref{sec:phot2} & 20.08 &   --- &  1.5 & --- \\
$u$    & 0.9      & 31.4  &  5538 &  0.15 & ---\\
$B$    & 0.95     & 31.4  &  2645 &  0.15 & 6501 \\
$V$    & 1.33     & 31.4  &  1350 &  0.15 & 5301 \\
$g$    & 1.58     & 31.4  &  1516 &  0.15 & 3601 \\
$r$    & 1.05     & 31.4  &  2166 &  0.15 & 5501 \\
$i$    & 0.95     & 31.4  &   799 &  0.15 & 3001 \\
$z$    & 1.15     & 31.4  &  1805 &  0.15 & 6401 \\
IA427  & 1.64     & 31.4  &   --- &  0.15 & 45001 \\
IA464  & 1.89     & 31.4  &   --- &  0.15 & 37001 \\
IA484  & 1.14     & 31.4  &   --- &  0.15 & 39601 \\
IA505  & 1.44     & 31.4  &   --- &  0.15 & 41701 \\
IA527  & 1.60     & 31.4  &   --- &  0.15 & 31301 \\
IA574  & 1.71     & 31.4  &   --- &  0.15 & 21701 \\
IA624  & 1.05     & 31.4  &   --- &  0.15 & 28501 \\
IA679  & 1.58     & 31.4  &   --- &  0.15 & 32201 \\
IA709  & 1.58     & 31.4  &   --- &  0.15 & 21901 \\
IA738  & 1.09     & 31.4  &   --- &  0.15 & 28601 \\
IA767  & 1.65     & 31.4  &   --- &  0.15 & 21401 \\
IA827  & 1.74     & 31.4  &   --- &  0.15 & 22701 \\
NB711  & 0.79     & 31.4  &  1200 &  0.15 & 141701 \\
NB816  & 1.51     & 31.4  &  5495 &  0.15 & 58301 \\
$Y$    & 0.9      & 30.0  & 24516 &  0.15 & ---\\
$J$    & 0.9      & 30.0  & 24516 &  0.15 & ---\\
$H$    & 0.9      & 30.0  & 24516 &  0.15 & ---\\
$K$    & 0.9      & 30.0  & 24516 &  0.15 & ---\\
$i1$   & See \S\ref{sec:phot2} & 21.58 &   --- &  0.6  & ---\\
$i2$   & See \S\ref{sec:phot2} & 21.58 &   --- &  0.6  & ---\\
$i3$   & See \S\ref{sec:phot2} & 21.58 &   --- &  0.6  & ---\\
$i4$   & See \S\ref{sec:phot2} & 21.58 &   --- &  0.6  & ---\\
$m24$  & See \S\ref{sec:phot2} & 20.15 &   --- &  1.2  & ---\\
$m70$  & See \S\ref{sec:phot2} & 17.53 &   --- &  4.0  & ---\\
$p100$ & See \S\ref{sec:phot2} &  8.9  &   --- &  1.2  & ---\\
$p160$ & See \S\ref{sec:phot2} &  8.9  &   --- &  2.4  & ---\\
$s250$ & 18.15    &  8.9  &   --- &  6.0  & ---\\
$s350$ & 25.15    &  8.9  &   --- &  8.3  & ---\\
$s500$ & 36.3     &  8.9  &   --- & 12.0  & ---\\
\hline
\end{tabular}
\label{tab:settings}
\end{center}
\end{minipage}
\end{table*}

\begin{table*}
\begin{minipage}{7in}
\begin{center}
\caption{Instrument-dependant \textsc{lambdar} settings}
\begin{tabular}{l|c|c|c|c|c|c|c|}
\hline
Instrument      & \textit{GALEX} & Optical, NIR & IRAC & MIPS 24 & MIPS 70 & PACS & SPIRE \\
\hline
PSFConvolve   & Yes        & No  & Yes      & Yes    & Yes      & Yes     & Yes \\
PixelFluxWgt  & Yes        & Yes & Yes      & Yes    & No       & Yes     & No \\
PSFWgt        & NUV only   & No  & Yes      & Yes    & Yes      & Yes     & Yes \\
Point sources & No         & No  & No       & Yes    & Yes      & Yes     & Yes \\
nIterations   & 5          & 2   & 5        & 5      & 5        & 5       & 5 \\
PSFConfidence & 0.95       & 1   & 0.95     & 0.95   & 0.95     & 1       & 1 \\
Prior         & $u < 24.0$ & --- & \S\ref{sec:phot2} & $i4 < 20.5$ & $m24 < 17.8$ & m24 detection & \S\ref{sec:phot2} \\
\hline
\end{tabular}
\label{tab:settings2}
\end{center}
\end{minipage}
\end{table*}

In order to obtain robust matched aperture photometry in the G10 region we ran \textsc{lambdar} to obtain photometry for all objects with $i < 25$~mag. A contaminant list is not required in the optical and near infrared because we are targeting all objects above a constant flux level. Again, we replaced all saturated regions in the Subaru imagery with a nominal value above the saturation threshold (see Table \ref{tab:settings}) equal to approximately 90\% of the maximum pixel value. This replacement also reduces artificial shredding of apertures. This causes somewhat incorrect photometry for bright, nearby and hence saturated objects, but is inconsequential because we are primarily focusing on fainter systems at intermediate redshifts. These objects are flagged in the individual band catalogues detailed in Section \ref{sec:release}.

For the IRAC and MIPS bands, the contaminant list consists of objects in the S-COSMOS IRAC, MIPS 24 and MIPS 70 catalogues that do not match to a source in our optically-selected sample with a radius of approximately half the PSF FWHM (1\arcsec, 3\arcsec and 9\arcsec respectively). We employ a similar technique in the far-infrared, matching against the PACS blind catalogues and HerMES StarFinder catalogues with a 3.9\arcsec, 6.0\arcsec, 9.0\arcsec, 12.6\arcsec and 18.2\arcsec radius in increasing wavelength order. In bands where PSF convolution is disabled, the minimum aperture radius was set to the PSF FWHM.

Figure \ref{fig:sky} suggests that there will be many objects that lie well below the detection threshold in bands with comparatively low resolution or sensitivity (i.e. the FUV, NUV and mid to far-IR bands). Attempting to obtain flux for systems significantly below the data sensitivity limits can be problematic. In particular the \textsc{lambdar} inbuilt flux-sharing will inevitably down-weight bright systems and up-weight the faint systems if flooded with targets given the poorer spatial resolution. Although \textsc{lambdar} has been designed to manage this at some level, eventually any algorithm will break-down if swamped with thousands of targets where only a few are detectable. Hence some prudent pruning of the optically-selected input and contamination catalogues is necessary for lower resolution and/or shallower data. Here we impose a set of cascading flux cuts and prior flux weights to prevent flux being scattered into these objects. Ultraviolet fluxes have prior weighting (using \textsc{lambdar}'s built-in functionality) derived from the \textsc{lambdar} $u$-band photometry, with any object fainter than $u > 24.0$~mag removed from the UV input catalogue. Fluxes in IRAC 1 and 2 were similarly weighted by the $K$ band flux with no cut applied to the target sources, but a $K < 23.5~$mag cut applied to the contaminant list. Fluxes in IRAC 3 and 4 were weighted by IRAC 2, with any source fainter than $i2 < 21.5$~mag and contaminant fainter than $i3 < 22.0$~mag removed. A similar cut of $i4 < 20.5$ was used in MIPS 24. Any source with MIPS 24 flux greater than zero was run in the PACS bands, and sources with $m24 < 19.0$, 18.0, 17.8 and 17.5 were run in SPIRE 250, 350, 500 and MIPS 70 respectively. Contaminants in these bands were weighted by their fluxes as measured in their corresponding catalogues. The IRAC flux cuts are illustrated in Figure \ref{fig:fluxcuts}.

To construct the photometric catalogue, we run \textsc{lambdar} with the settings shown in Tables \ref{tab:settings} and \ref{tab:settings2}. While measuring flux, we apply multiplicative aperture corrections of 1.15 for MIPS 70 \citep{frayer09}, 1.50 for PACS 100 and 1.477 for PACS 160, and an additional multiplicative high-pass correction of 1.12 and 1.11 for PACS 100 and 160 respectively\footnote{\url{http://www.mpe.mpg.de/resources/PEP/DR1_tarballs/readme_PEP_global.pdf}}.


For PSF convolution, \textsc{lambdar} accepts either a FWHM assuming a Gaussian shape or an empirical PSF provided in a FITS file. We use the \citet{hora12} PSF convolution kernels in IRAC 1 and 2, the \citet{gordon08} kernels in the remaining IRAC and MIPS bands (100~K for MIPS) and the observed PSF kernels from PEP. \textit{GALEX} convolution kernels were provided by the \textit{GALEX}-GAMA team. For other bands, we use a Gaussian PSF with FWHM given in Table \ref{tab:settings}.

For MIPS 24 and longer wavelengths, we obtain point source photometry only. This reduces the problem of optically defined large apertures (e.g. of an elliptical galaxy) being no longer appropriate because of the significantly decreased sensitivity in MIPS 24 (see Figure \ref{fig:sky}) and declining SEDs. The flux missed will be minimal even in extended sources for the same reasons. Finally, we reconstruct errors in all bands to be the sum of the deblend error, sky rms and sky flux error in quadrature. This omits a shot noise term that was unrealistically large in the \textit{GALEX} and UltraVISTA bands (due to the GAIN assumed by \textsc{lambdar}) and dominated the error budget for the overwhelming majority of objects. 

\subsection{Star-galaxy separation}

In order to robustly identify stars in our photometric catalogue, we perform a multiple stage star-galaxy separation process. Firstly, we apply the star-galaxy flags derived by the COSMOS2015 team. Full details of this process are described in \citet{laigle16}. Briefly, sources are fit using \textsc{Le Phare} for both galactic and stellar templates, and best fits derived for both. A source is classed as a star if i) its best fit $\chi^{2}_{gal}-\chi^{2}_{star}>0$, ii) it is detected in the NIR or IRAC bands, and iii) it lies close to the stellar sequence in $BzK$ colour space. In addition, the COSMOS2015 catalogue contains a flag to indicate the source is x-ray detected (and potentially an AGN), which we also propagate to our final catalogue.

Following this we then perform our own stellar identification using the size-magnitude distribution given in Figure \ref{fig:magsize-after}. We identify stars are sources with semi-major axis $< 4^{\prime\prime}$, semi-major-axis - $i <$ 22 and $i<$ 21 (where $i$ denotes the final \textsc{lambdar} magnitude), displayed in Figure \ref{fig:magsize-after} as the polygon in the bottom left hand corner bounded by the orange and blue lines. The stellar classification from our size-magnitude distribution only supersedes sources which are classified as galaxies by COSMOS2015. For sources which do not have a star-galaxy flag in the COSMOS2015 catalogue (predominately new sources added in our selection), we apply our size-magnitude flags.

As a final stage, we visually inspect sources in two subsamples and class objects as either a star or galaxy. For all visual inspections we use cutouts from the HST-F814W data which allows the most robust star-galaxy separation. Firstly, we inspect all $\sim15,000$ sources which are classed as either a star or an x-ray source following the COSMOS2015 and size-magnitude assignments given above. This process identifies any galaxies which have been falsely assigned as stars using our previous selection. Secondly, we visually inspect all sources which have not been identified as stars and are either at $z<0.06$ (a key epoch in future studies involving this dataset) or with $i$-mag$<$22 and with no secure spectroscopic redshift (such sources potentially have erroneously bright magnitudes). In total this sample contains $\sim5,000$ sources, of which 385 are identified as stars. These visual classifications then supersede the previous classifications, resulting in a ‘Master’ star-galaxy separation flag using information from COSMOS2015, our size-magnitude classification and visual inspection, where $\sim7$ per cent of sources in our catalogue are classed as stars, $\sim92$ per cent are classed as galaxies and $\sim1$ per cent as x-ray sources. In total there are 5539 sources which are visually identified as galaxies, 744 resolved x-ray sources and 831 x-ray point sources.

All individual star-galaxy flags and our resultant ‘Master’ star-galaxy flag are given in our publicly available photometric catalogue.

\section{Consistency checks}
\label{sec:con}

\subsection{Astrometry}
\label{sec:astrometry}

\begin{figure*}
\begin{minipage}{7in}
\begin{center}
\includegraphics[width=0.9\linewidth]{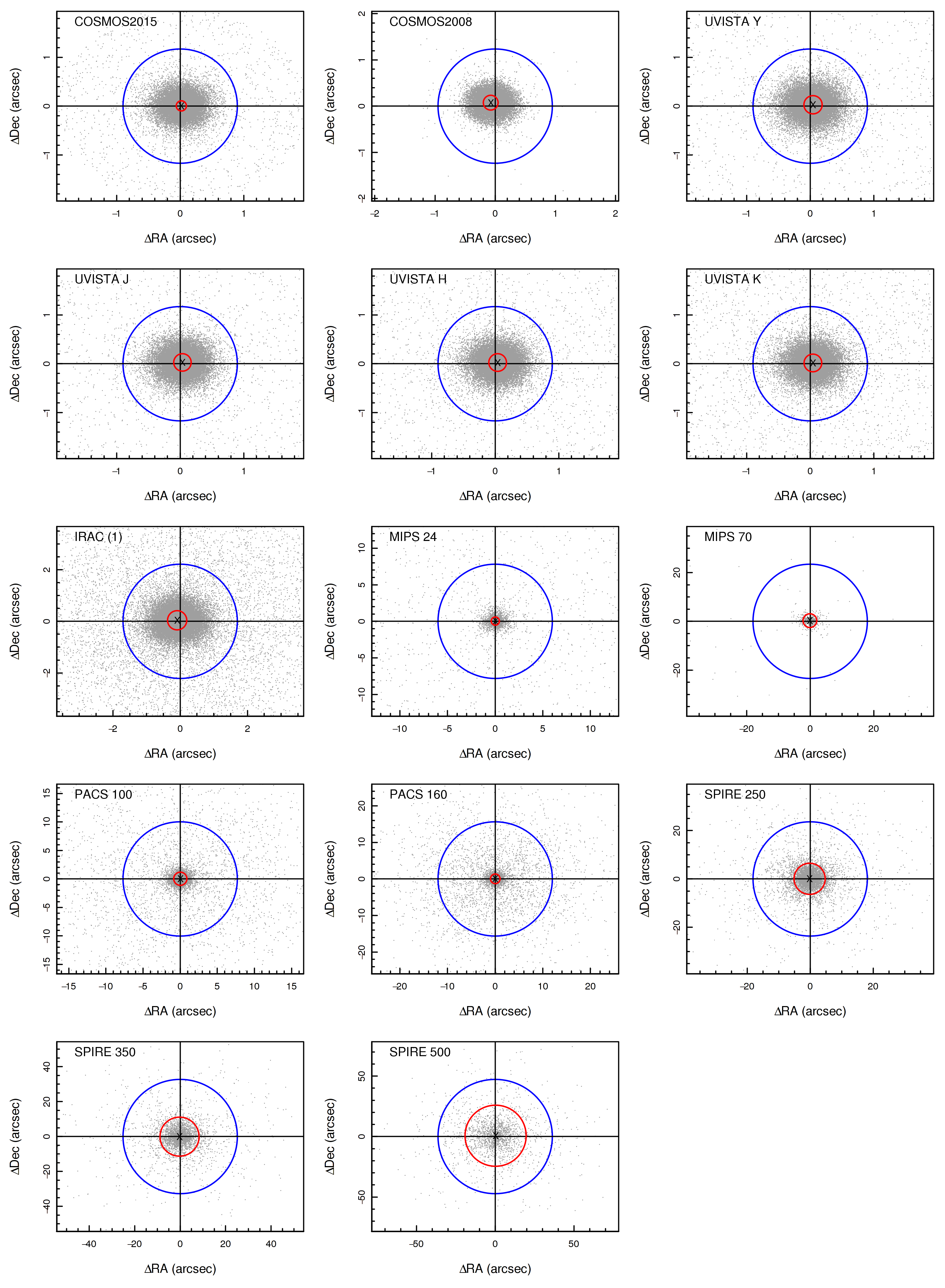}
\caption{Positional offsets between the aperture catalogue and various archive catalogues. The blue circle shows the PSF FWHM, the black cross the median positional offset and the red circle contains 80\% of the population after taking into account random mismatches. When multiple bands are included in the comparison catalogue, the band used is indicated in brackets.}
\label{fig:astrometry}
\end{center}
\end{minipage}
\end{figure*}

Figure \ref{fig:astrometry} shows the positional offset between our final aperture catalogue and the pre-existing archival data. The catalogues used for this comparison are COSMOS2015, the 2008 update to the \citet{capak07} catalogue (providing photometric measurements for \textit{GALEX}, CFHT and Subaru), the UltraVISTA DR2 basic blind catalogues, the S-COSMOS IRAC, MIPS 24 and MIPS 70 catalogues, the PEP DR1 blind catalogues and the HerMES DR2 StarFinder catalogues. For this analysis, positional matches were performed with a radius of three times the PSF FWHM except for COSMOS2015 and COSMOS2008, where exact ID matches were used. 

The centroids in the Figure, depicted by the black cross and computed after removing random mismatches from the sample, demonstrate that our astrometry is robust to within 0.12 times the PSF FWHM. For the optical data, this translates to an astrometric accuracy of 0.106\arcsec. In all other cases, the centroid is within 0.1\arcsec, apart from the low-resolution far-infrared bands (MIPS 70 and SPIRE), where it remains below 0.75\arcsec. 66 per cent of the data, as delineated by the red circle, is contained with 0.2 times the PSF FWHM at all wavelengths except for the SPIRE bands. Sources mismatching outside the PSF (as denoted by the blue circle) are generally small extended objects (1--2\arcsec). From this, we conclude that our astrometric accuracy is as expected given the PSF FWHM, apart from potentially SPIRE 500 where the resolution and sensitivity are the lowest.


\subsection{Comparisons with existing photometry}
\label{sec:trumpets}

\begin{figure*}
\begin{minipage}{7in}
\begin{center}
\includegraphics[width=1.0\linewidth]{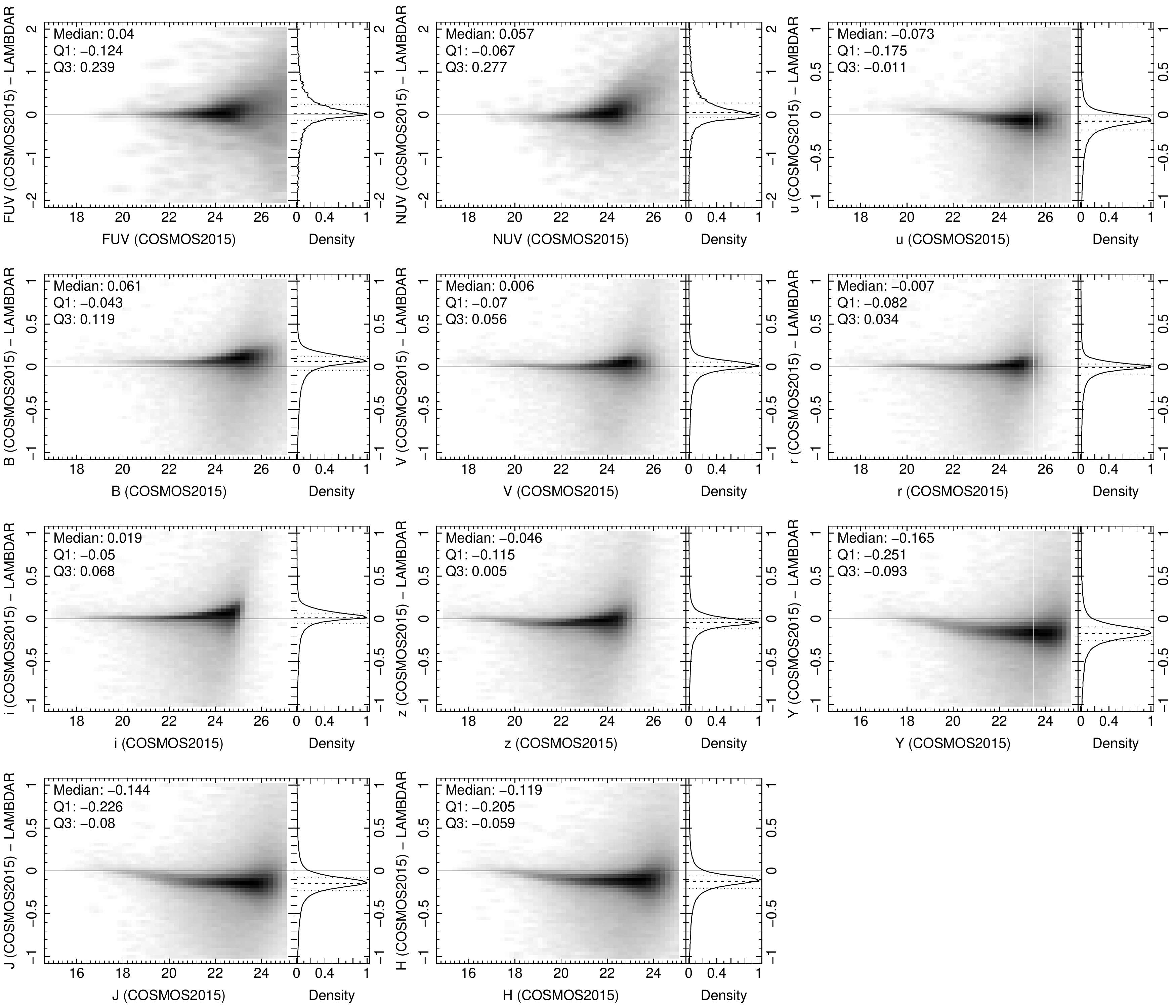}
\caption{Magnitude offsets between the photometry derived in this work and COSMOS2015; this is a 2D kernel density estimation with square root scaling. The histograms on the right show the 1D distribution and give the median and interquartile range, which are also inset in the main panels.}
\label{fig:trumpets}
\end{center}
\end{minipage}
\end{figure*}

\begin{figure*}
\begin{minipage}{7in}
\begin{center}
\includegraphics[width=1.0\linewidth]{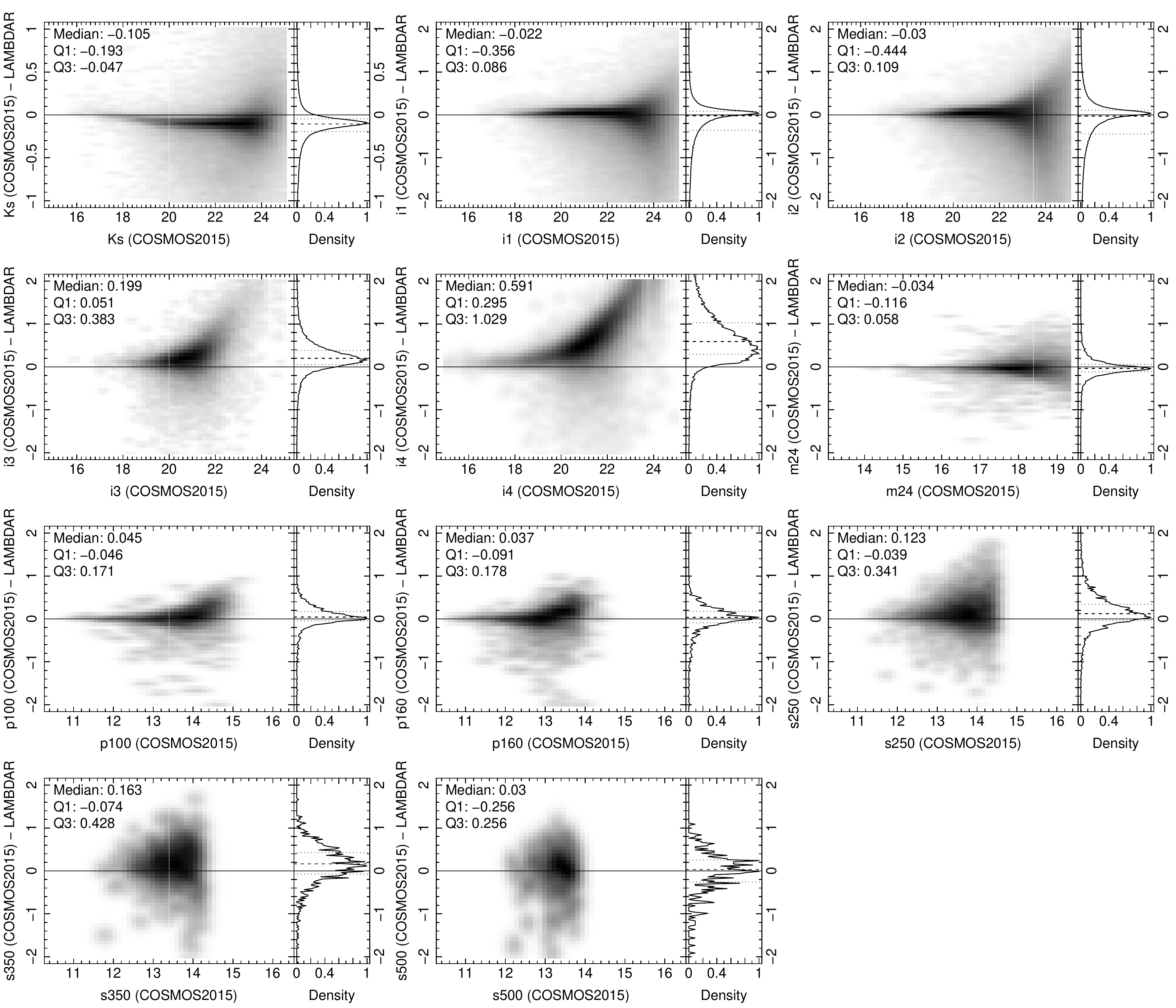}
\contcaption{}
\end{center}
\end{minipage}
\end{figure*}

Figure \ref{fig:trumpets} compares magnitudes derived from \textsc{lambdar} and the COSMOS2015 catalogue \citep{laigle16} (AUTO magnitudes). For these comparisons, saturated and masked objects are removed. An exact match for the 2008 COSMOS ID was performed. A value above zero in this Figure indicates \textsc{lambdar} recovers more flux than the comparison catalogue and vice versa.

Our photometric measurements are broadly consistent with the COSMOS2015 photometry, with two exceptions. In the near infrared, our catalogue is consistent with the UltraVISTA DR2 blind detections. The offset from COSMOS2015 may be due to differences in aperture definition or choice of selection image. In IRAC 4, our catalogue is consistent with S-COSMOS. The photometry for the Subaru narrow bands, while not shown in Figure \ref{fig:trumpets}, is also consistent with the COSMOS2015 catalogue. While not apparent in Figure \ref{fig:trumpets}, a population of objects with $24.5 < i < 25$ may have apertures that are an erroneous combination of objects near or below $i = 25$~mag. This arises from the hard flux cut being made in the \textsc{lambdar} input catalogue. For similar reasons, our catalogue contains 18323 objects not in COSMOS2015. Therefore, fluxes for objects with $i > 24.5$~mag or not in COSMOS2015 should be treated with caution.

\subsection{Colour distributions}

\begin{figure*}
\begin{minipage}{7in}
\begin{center}
\includegraphics[width=1.0\linewidth]{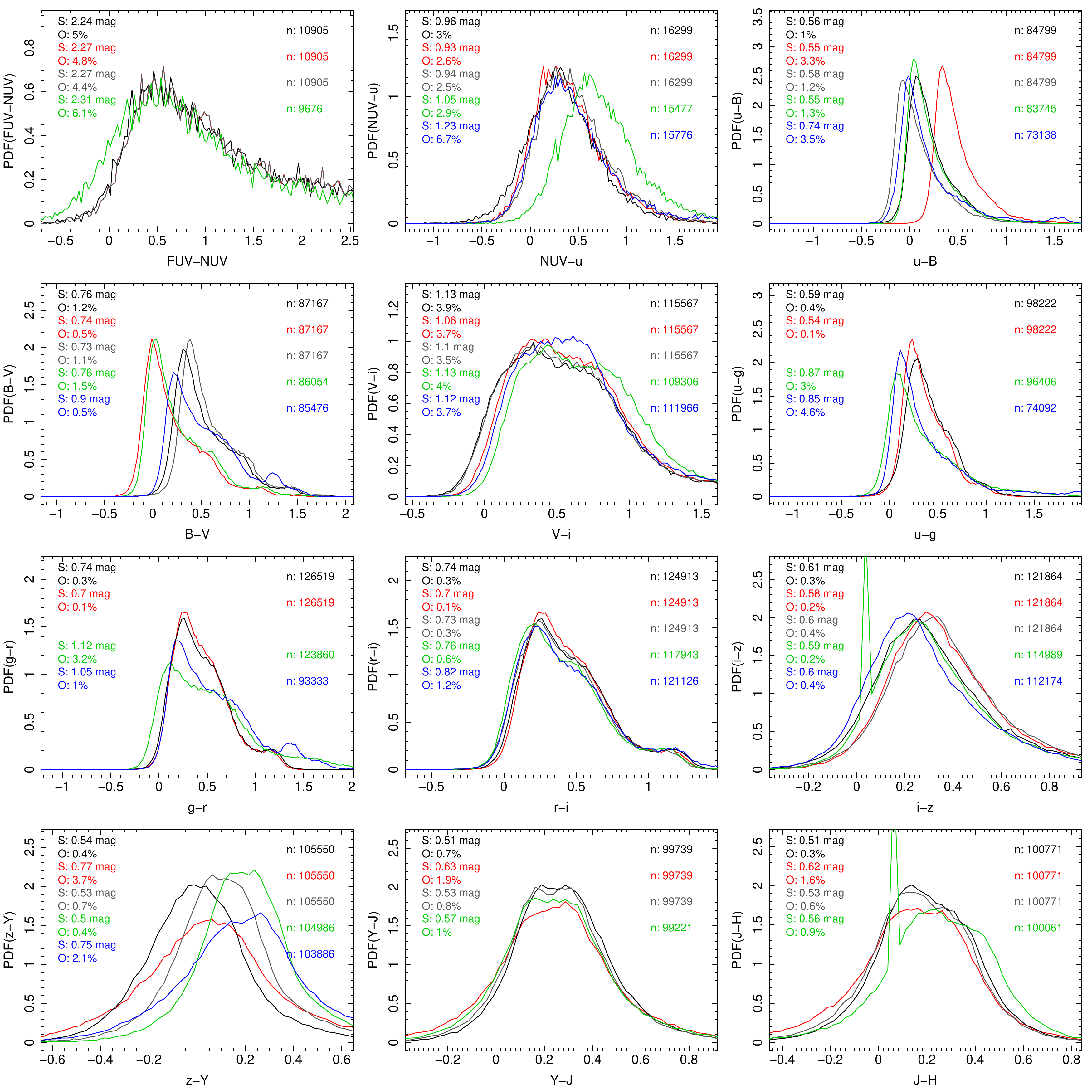}
\caption{PDFs of colour distributions for adjacent filters (as indicated) for the \textsc{lambdar} (black), archival (red), \citet{laigle16} (grey), \citet{muzzin14} (green) and CFHT-LS (blue) photometry. The 80th percentile spread and outlier rate (objects that lie outside 0.5~mag of the 80th percentile range) are denoted by S and O respectively. A flux cut of 0.5~mag fainter than the peak of the number counts in each band of our catalogue was applied to all data sets.}
\label{fig:colours}
\end{center}
\end{minipage}
\end{figure*}

\begin{figure*}
\begin{minipage}{7in}
\begin{center}
\includegraphics[width=1.0\linewidth]{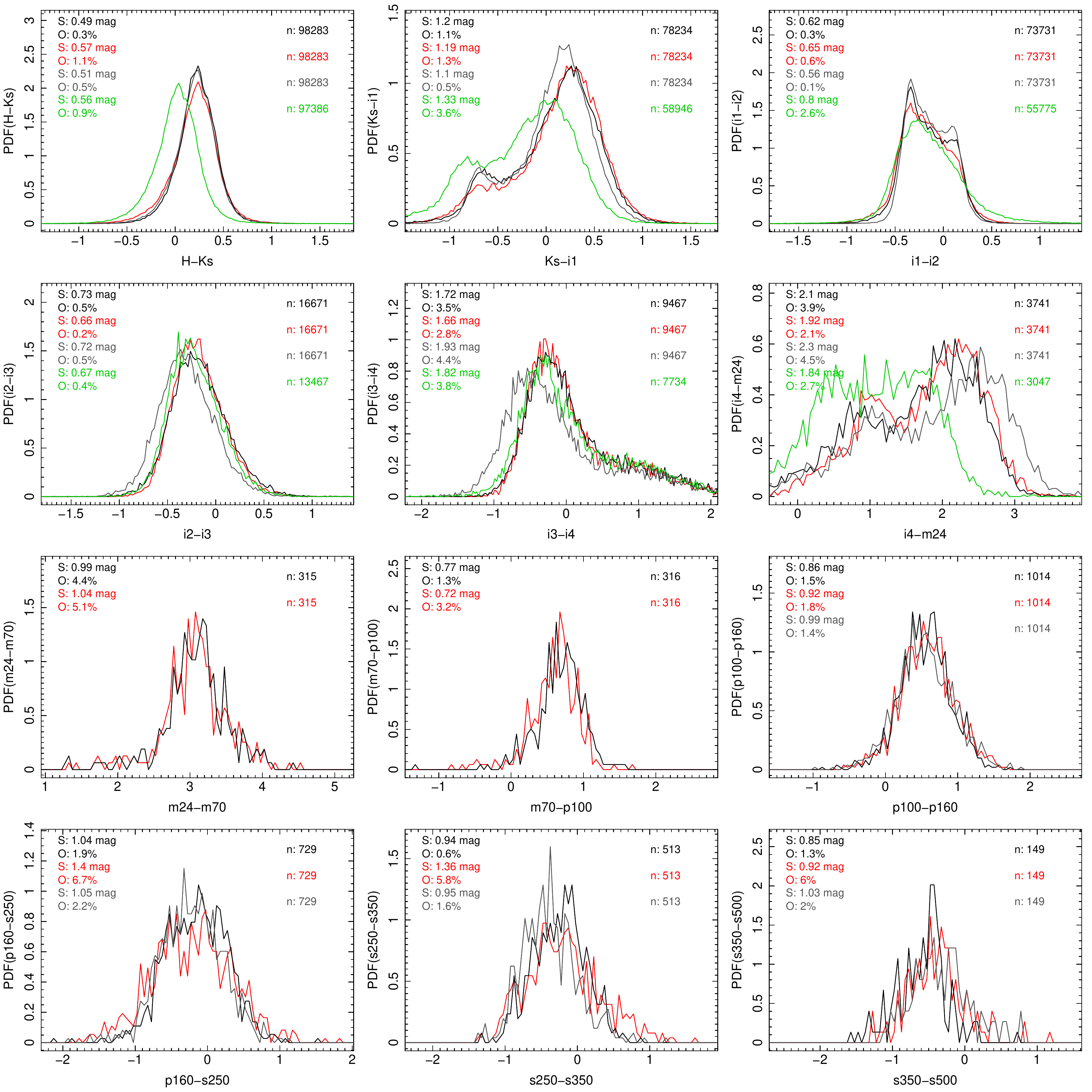}
\contcaption{}
\end{center}
\end{minipage}
\end{figure*}

The plots shown in Figure \ref{fig:trumpets} are useful for diagnosing zeropoint and linearity problems, but do not give an objective assessment as to which data set is more robust. In order to shed light on this, we examine distributions of adjacent colours across all broad bands. A sample of galaxies will have some intrinsic colour distribution, which is then convolved with error distributions introduced by the instrumentation, observing conditions, photometric data reduction methods and photometric measurement error. A narrower colour distribution and a lower outlier fraction --- indicating a narrower photometric error distribution if the same images and galaxies are used --- are, hence, more desirable. 

To this extent, we compare our colours to those derived from the archival photometry detailed in Section \ref{sec:astrometry}, COSMOS2015, \citet{muzzin14} ($r$ band selected) and CFHT-Legacy Survey (CFHT-LS) in Figure \ref{fig:colours}. In this comparison, we use the COSMOS2015 AUTO magnitudes. \citet{muzzin14} derived PSF matched 2.1\arcsec photometry based on the publicly available GALEX, CFHT, Subaru, UltraVISTA, S-COSMOS IRAC and MIPS 24 imagery detailed in Section \ref{sec:data}. CFHT-LS independently surveyed a 1 deg$^2$ portion of the COSMOS region in the MegaPrime $ugrizy$ filters, achieving 80 per cent completeness at $i = 25.5$~mag. The corresponding CFHT-LS catalogue contains AUTO magnitudes, which are converted to the Subaru filterset via the equations in \citet{capak07}. 

For these comparisons, catalogues were matched with the radii given in Section \ref{sec:trumpets} and saturated and masked objects are removed. This is a matched sample, meaning the colour distributions reflect only sources that are in our catalogue, the archival catalogue and the COSMOS2015 catalogue. The red, green, and blue curves show the subset of the matched sample that have \citet{muzzin14} and CFHT-LS photometric measurements. As the imaging data used is common to all sets apart from CFHT-LS, it is only the photometric measurement that is being tested.  

To indicate the width of the colour distribution, we somewhat arbitrarily use the 80th percentile spread. To determine the rate of gross photometric errors, we define outliers to be objects that lie 0.5~mag outside the 80th percentile of each colour distribution. It is plausible that objects have intrinsic colours that fall within this region, especially when the underlying colour distribution and/or redshift range are broad. This is particularly noticeable at optical wavelengths, where the broad distribution of rest-frame FUV, NUV and $u$ band emission arising from the wide variance between star forming and passive systems has been redshifted into $B$, $V$, $g$ and $r$ and superimposed on lower redshift objects. However in the near-infrared where the intrinsic spread is the lowest, the outlier rate almost purely reflects gross phometric errors.

We find that our catalogue is comparable to the COSMOS2015 AUTO colours in nearly all bands. Our catalogue is also comparable to the other, existing catalogues, while avoiding the spikes in $i-z$ and $J-H$ and the colour offsets in $K-i1$ and $i4-m24$ seen in the \citet{muzzin14} catalogue. The inconsistency between the datasets for the Subaru $B$ filter is due to the zeropoint calibration being uncertain at the time of the 2008 COSMOS photometric catalogue.

The boundaries between the different data reduction techniques (NUV$-u$, $z-Y$, $K_s-i1$, $i4-m24$, $m24-m70$, $m70-p100$, $p160-s250$) are also of interest because they highlight the non-homogenous nature of the existing photometry, which manifests in colour offsets between the different datasets. These boundaries also exhibit increased outlier rates due to the potential for table mismatches and zeropoint errors.

Our catalogue generally has comparable errors to the archival photometry in the optical, MIPS and PACS passbands, and \textit{GALEX}. The use of \textsc{lambdar} to derive consistent errors across the wavelength range should represent a significant improvement over table matching archival catalogues at least three further aspects. Firstly, no aperture corrections are provided for the errors in the archival catalogues and a simple scaling is likely to underestimate the expanded error. Secondly, the use of \textsc{SExtractor} to derive errors in the archival UltraVISTA and IRAC catalogues may lead to errors being systematically underestimated (see \citealt{hill11}). Finally, \textsc{lambdar} incorporates a (conservative) estimate of the uncertainty in deblend solutions, leading to more realistic errors for objects in crowded regions.

\subsection{SED fitting}

\begin{landscape}
\begin{figure}
\begin{minipage}{9in}
\includegraphics[width=0.23\linewidth]{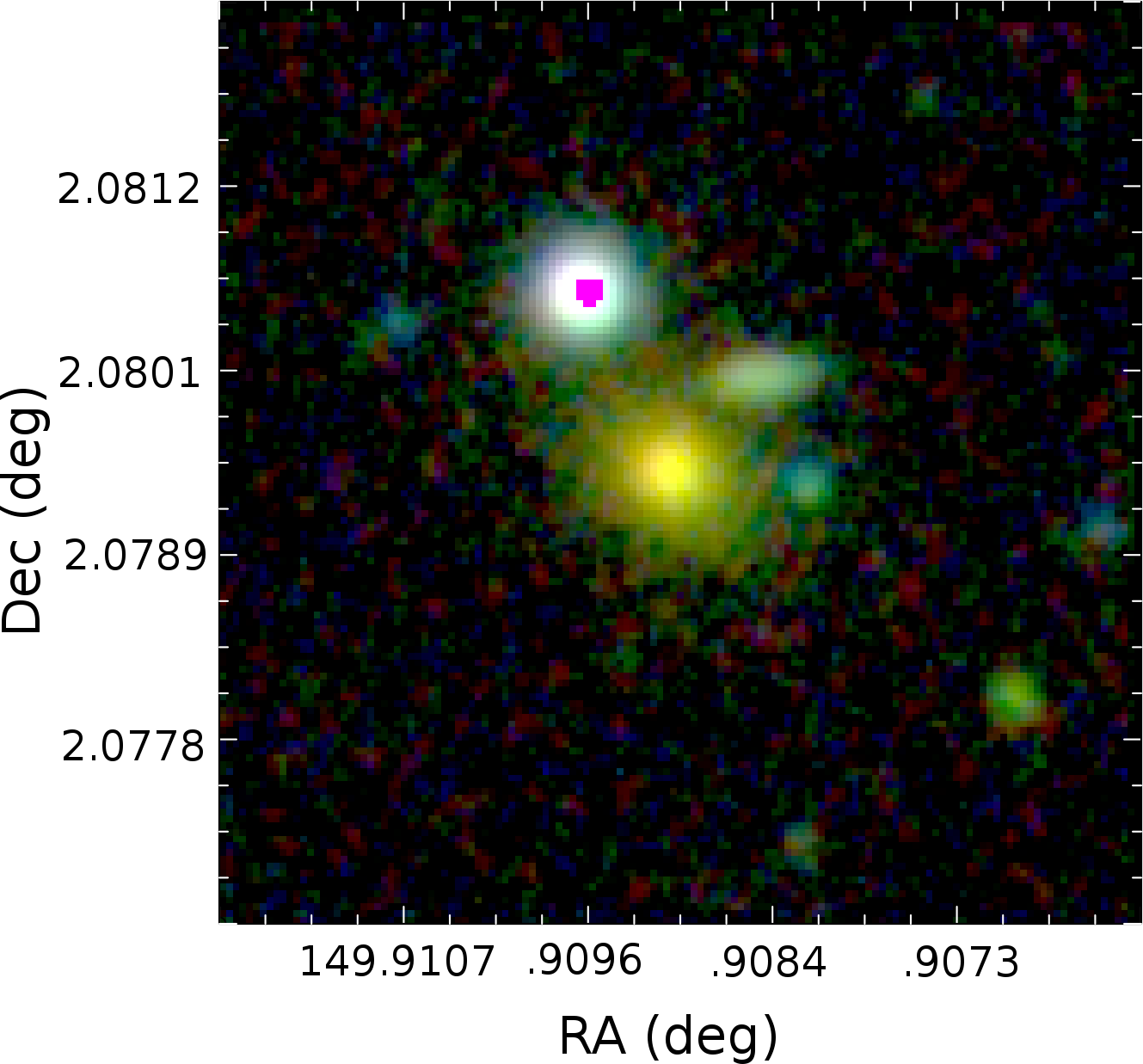}
\includegraphics[width=0.38\linewidth]{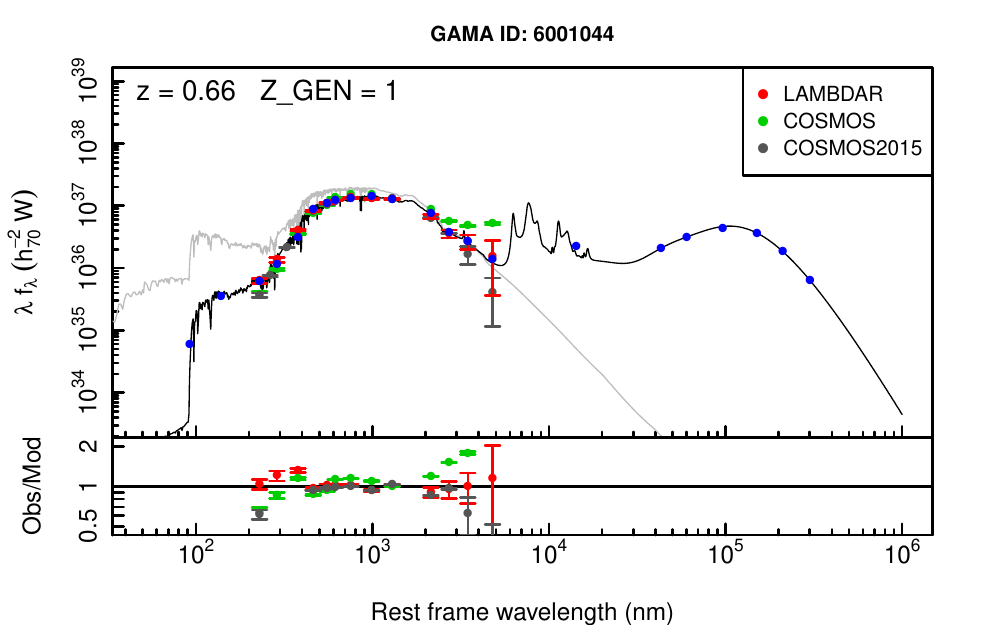}
\includegraphics[width=0.38\linewidth]{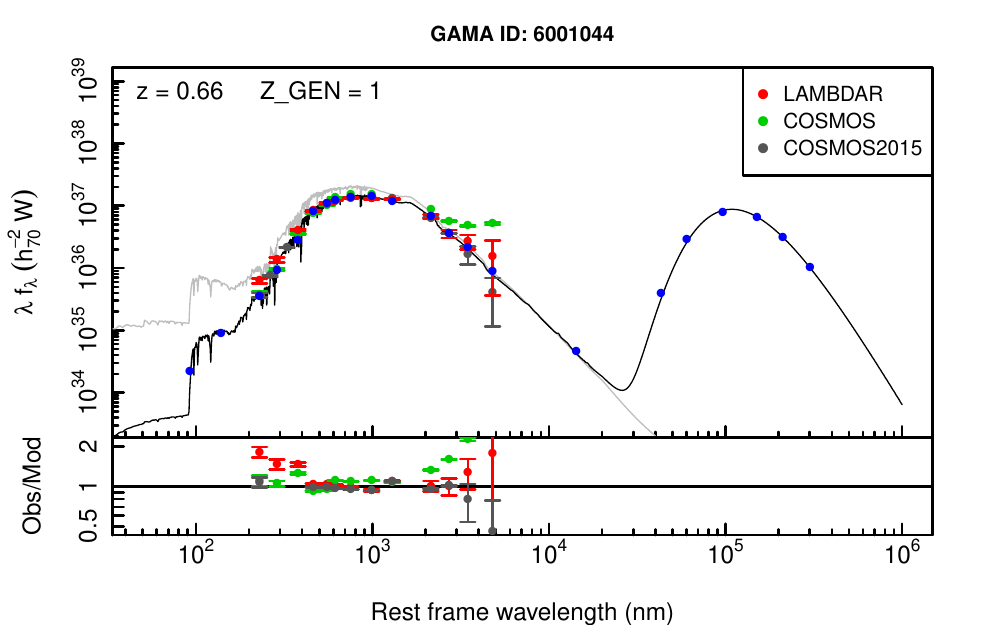} \\

\includegraphics[width=0.23\linewidth]{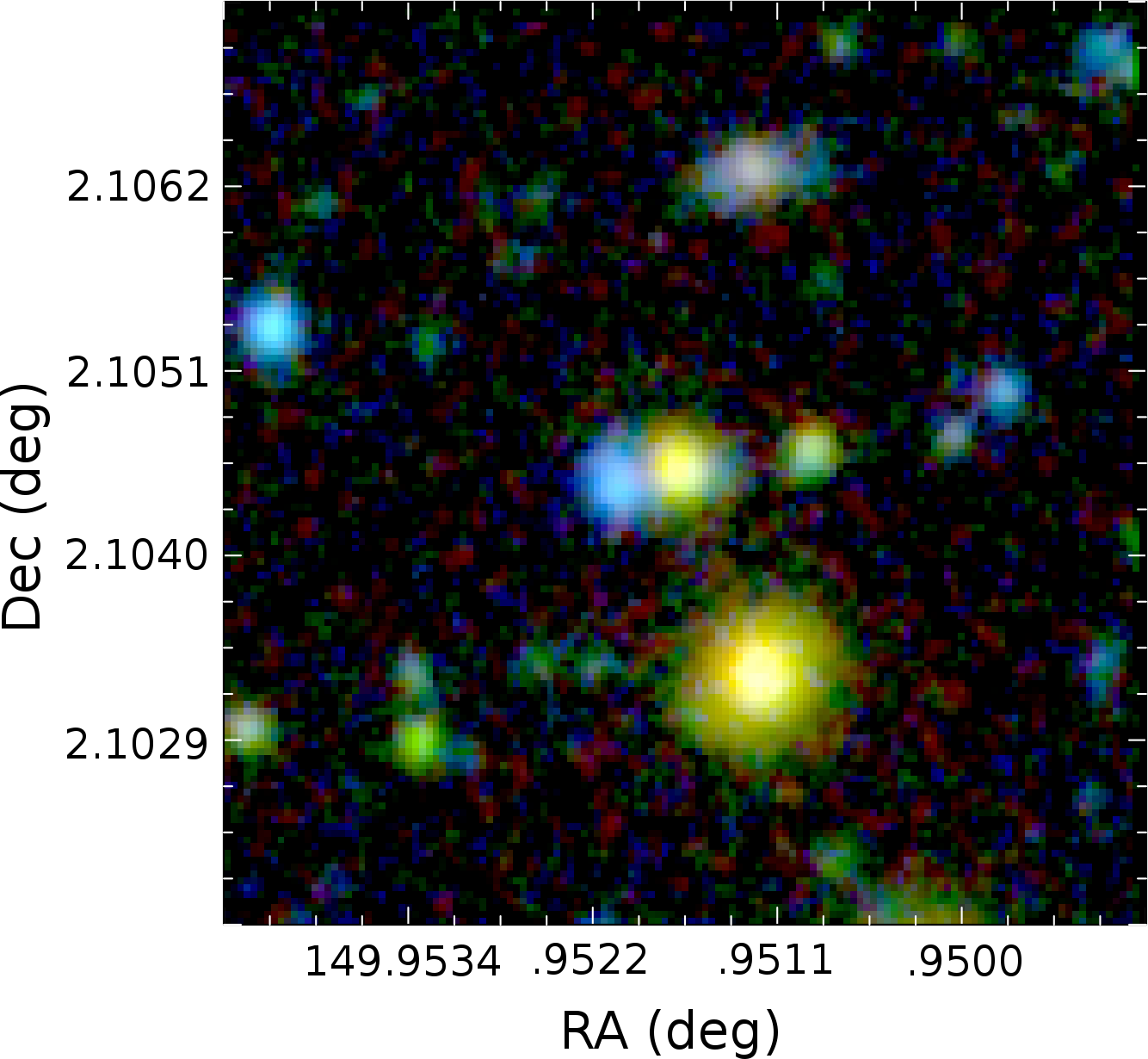}
\includegraphics[width=0.38\linewidth]{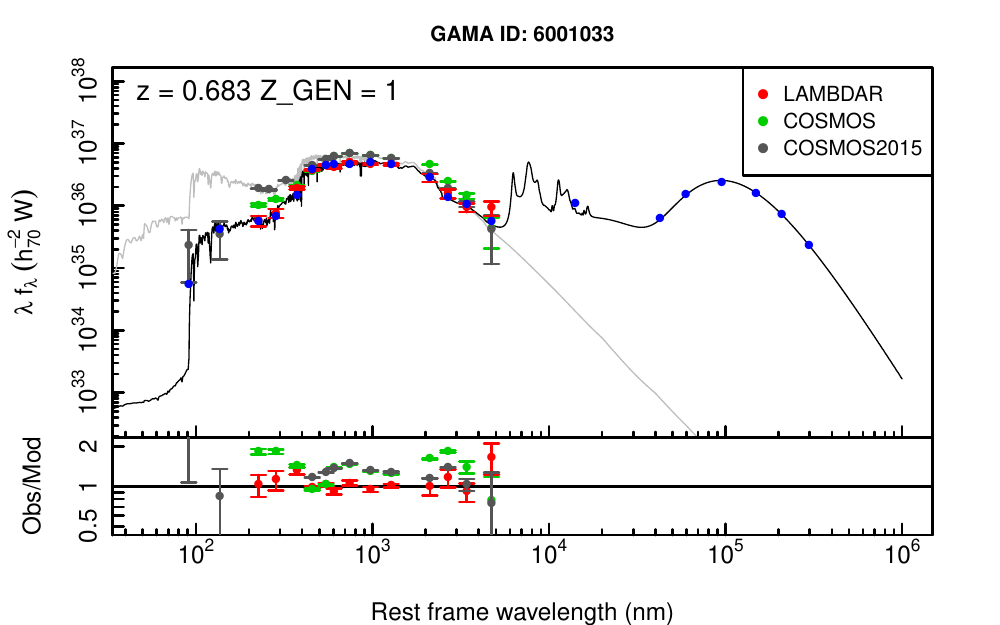}
\includegraphics[width=0.38\linewidth]{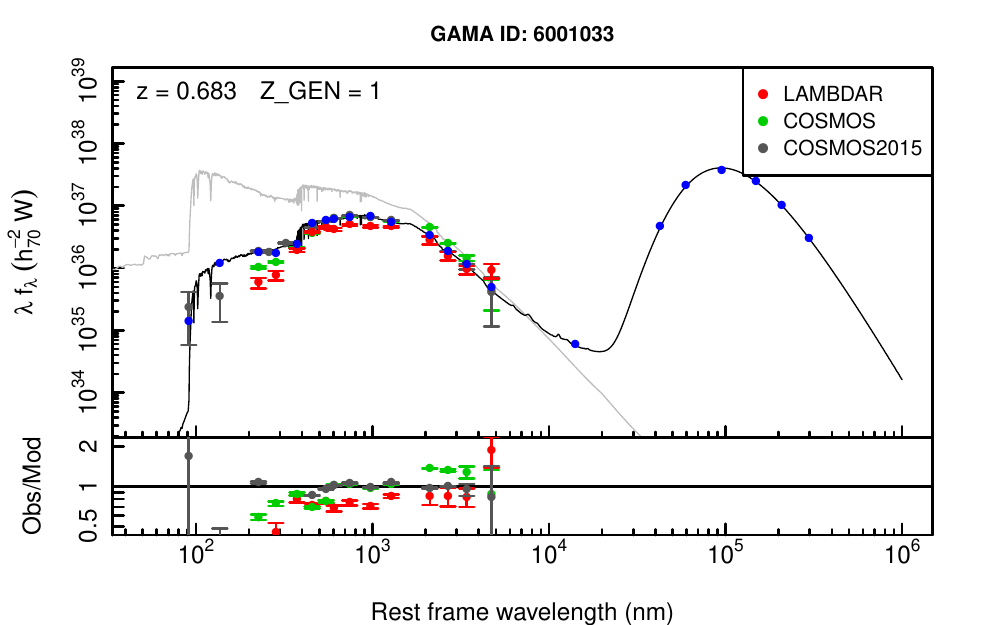} \\

\includegraphics[width=0.23\linewidth]{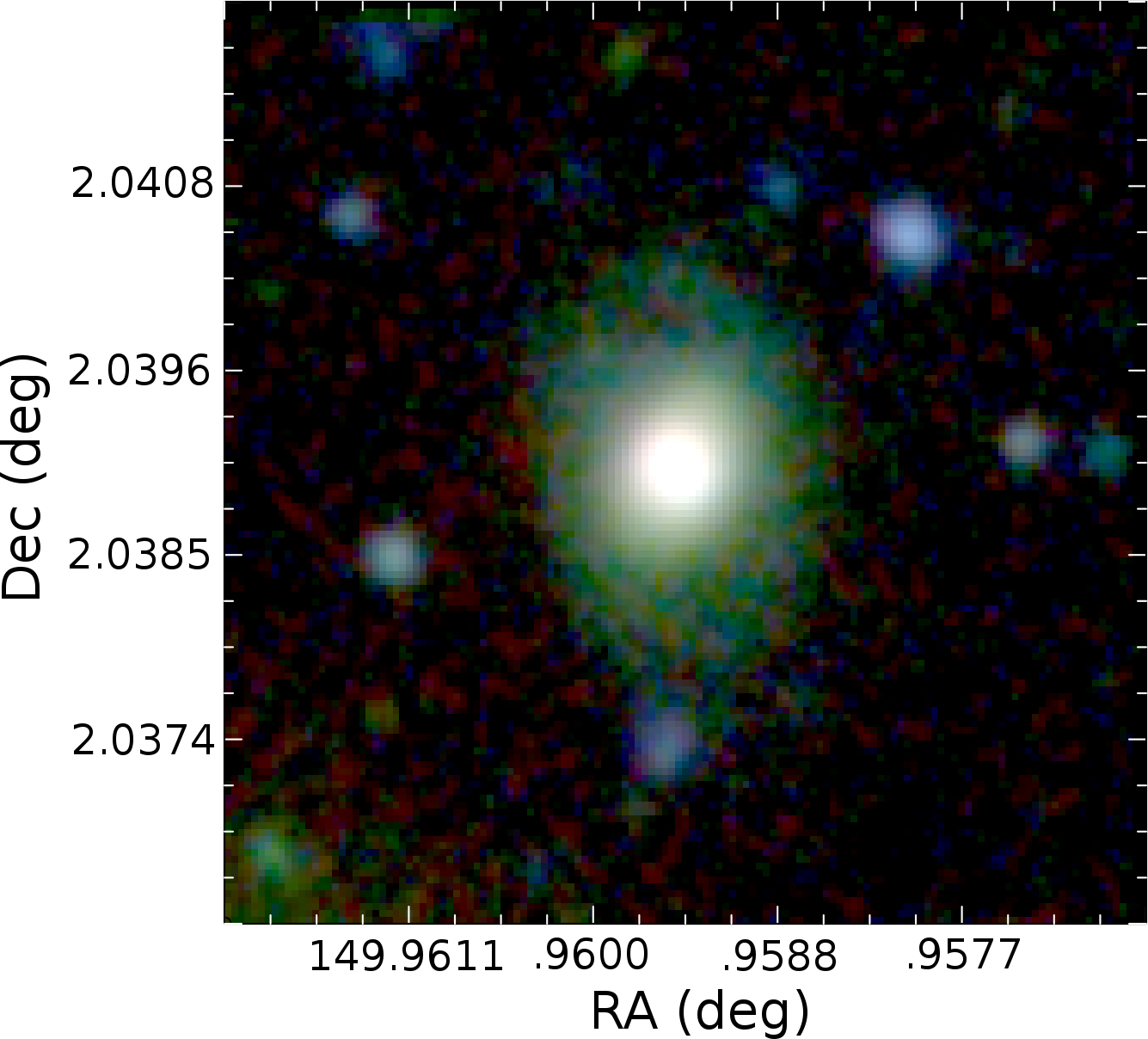}
\includegraphics[width=0.38\linewidth]{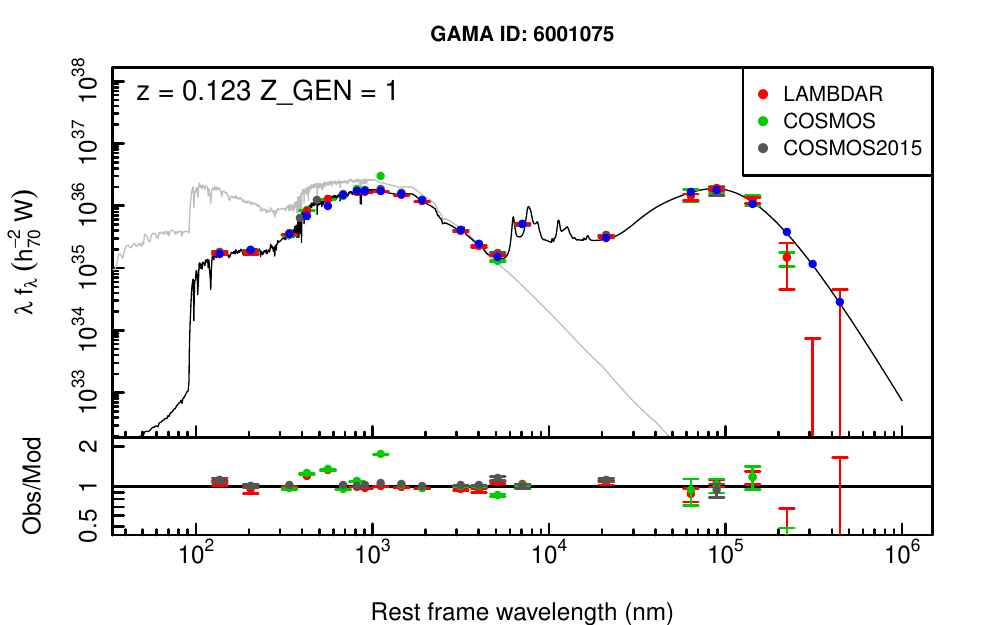}
\includegraphics[width=0.38\linewidth]{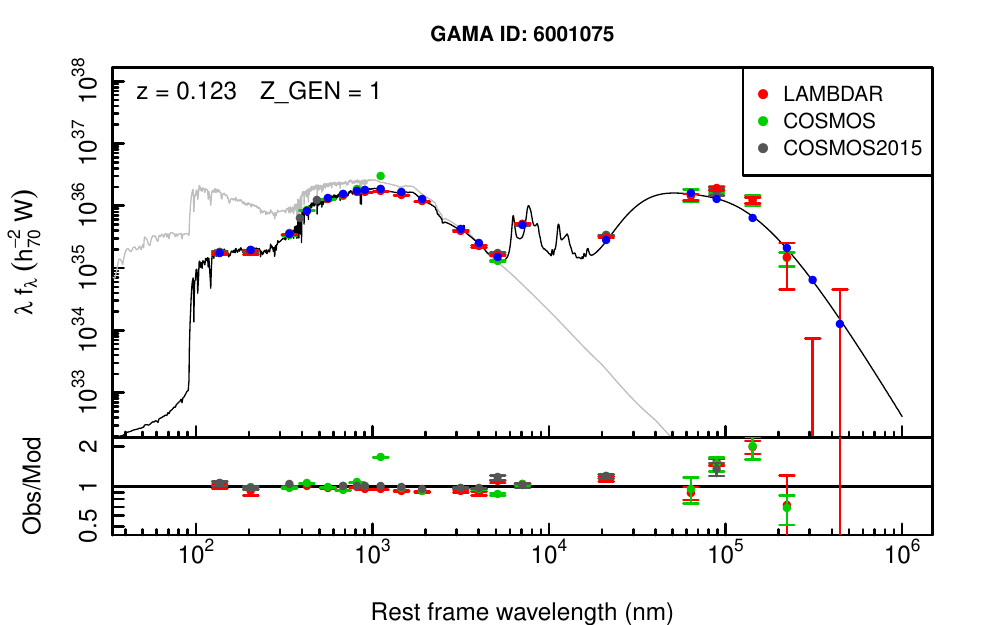} \\

\caption{Top: $Kiu$ cutouts for IDs 6001044 (top), 6001033 (centre) and 6001075 (bottom). Other panels: \textsc{magphys} fits for using the \textsc{lambdar} photometric catalogue produced in this work (center), the same but using COSMOS2015 photometry (right) for IDs 6001044 (top) and 6001033 (centre) and archival photometry for 6001075 (bottom). Black line: attenuated SED, grey line: unattenuated SED, blue points: attenuated SED convolved with the respective filters, green points: archival photometry, grey points: COSMOS2015 photometry, red points: our photometric catalogue. Z\_GEN, from \citet{davies15}, denotes the genesis of the redshift measurement; the three examples presented above have secure spectroscopic redshifts from zCOSMOS data.}
\label{fig:magphys}
\end{minipage}
\end{figure}
\end{landscape}

\begin{figure}
\begin{center}
\includegraphics[width=0.99\linewidth]{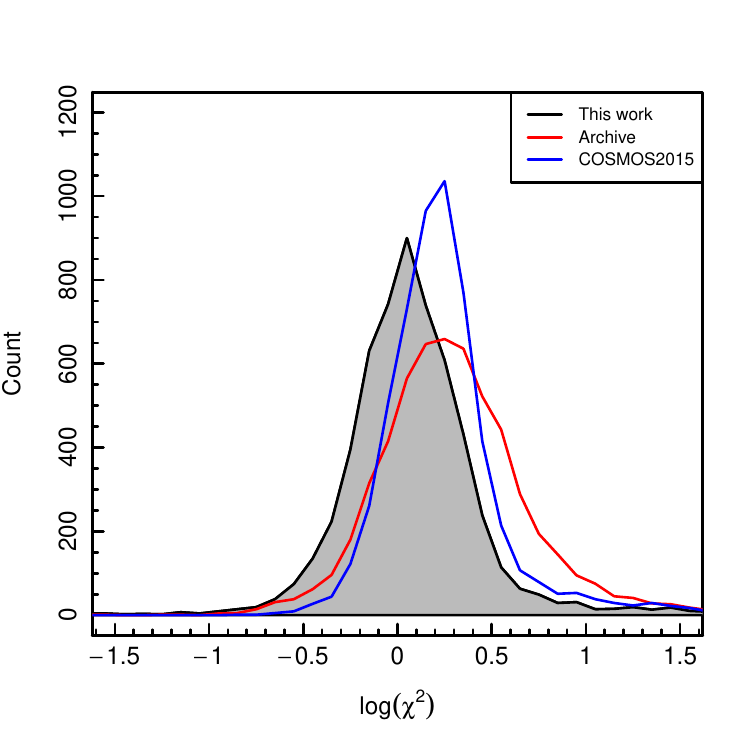}
\caption{The log($\chi2$) distribution of \textsc{magphys} fits using both our catalogue, the archival photometry and COSMOS2015 for approximately 5600 sources. Bin size is 0.1 in log($\chi2$).}
\label{fig:chi2}
\end{center}
\end{figure}

To highlight the benefits in producing a consistent total flux catalogue for full SED analysis and as a prelude to future investigations, we fit SEDs to 5619 sources (IDs 6000000 to 6006500) using \textsc{magphys} (Multi-wavelength Analysis of Galaxy Physical Properties; \citealt{dacunha08}) for both our catalogue, COSMOS2015 (AUTO magnitudes) and the archival photometry detailed in Section \ref{sec:astrometry}. The sample size was selected to give a quick look into the dataset while keeping CPU and manual inspection time requirements at a manageable level. Of these sources, $\sim98$ per cent have high resolution spectroscopic redshifts. We look to use \textsc{magphys} as a means of providing a representative interpolation function for the various photometric measurements. Uniform broad band coverage across a wide wavelength range is more important for SED fitting than the density of sampling within that range. In our fits, we use the wideband filters excluding $B$ and $V$, except for COSMOS2015 which uses these filters instead of $g$ and $r$.

Figure \ref{fig:magphys} shows example fits using our catalogue (left) and COSMOS2015 or the archival photometry (right). The top panel shows a situation where the deblending of \textsc{lambdar} is likely to be superior. The centre panel shows 6001033, which is deblended into two objects in our catalogue and is one object in COSMOS2015. Similarly, a clear break occurs between $z$ and $Y$ in the archival photometry for 6001033 due to different deblend solutions and methods. The erroneously high $Y$ band measurement for 6001075 in the archival catalogue is due to a large, incorrect aperture that loops around a nearby bright source. In this case, the COSMOS2015 fit is similar to the \textsc{lambdar} fit, with a noticeable difference in the far-infrared. 

The $\chi^2$ distribution for all three data sets is shown in Figure \ref{fig:chi2}. Broadly speaking, our catalogue achieves a $\chi^2$ distribution that is more balanced and centred on the expected $\chi^2 =1$ compared to both COSMOS2015 and the archive datasets. 3174 of the 5579 common objects have a $\chi^2$ closer to one in our catalogue compared to COSMOS2015. Note that the COSMOS2015 catalogue is optimised for accurate colours, while we aim to measure total fluxes. Compared to the archive, 3704 of the 5619 common objects have a $\chi^2$ closer to one with our catalogue. The improvement in $\chi^2$ is partly due to the incorporation of deblend errors into our catalogue, particularly for the IRAC bands. However, with a visual inspection it becomes clear that at least some of the improvement over the archive is due to the use of consistent photometry and wavelength errors across the wavelength range. 

These comparisons are inevitably complicated by the ability of \textsc{magphys} to correctly model galaxies at intermediate redshifts. This will be discussed in more detail in Driver et al. in prep, where we fit the entire GAMA and G10/COSMOS samples using both this and the \citet{wright16a} catalogues. Using these datasets, we will examine the cosmic spectral energy distribution (Andrews et al. in prep), star formation rates (Davies et al. in prep), stellar masses (Wright et al. in prep) and validate and improve the SED fitting process out to $z = 1$.

\section{Release content}
\label{sec:release}

\begin{table}
\caption{Galactic extinction corrections}
\begin{tabular}{l|c|c|}
\hline
Band  & $k(\lambda)$ \\
\hline
FUV   & 8.376$^a$ \\
NUV   & 8.741$^a$ \\
$u$   & 4.690$^b$ \\
$B$   & 4.039$^b$ \\
$V$   & 3.147$^b$ \\
$g$   & 3.738$^b$ \\
$r$   & 2.586$^b$ \\
$i$   & 1.923$^b$ \\
$z$   & 1.436$^b$ \\
IA427 & 4.260$^c$ \\
IA464 & 3.843$^c$ \\
IA484 & 3.621$^c$ \\
IA505 & 3.425$^c$ \\
IA527 & 3.264$^c$ \\
IA574 & 2.937$^c$ \\
IA624 & 2.694$^c$ \\
IA679 & 2.430$^c$ \\
IA709 & 2.289$^c$ \\
IA738 & 2.150$^c$ \\
IA767 & 1.996$^c$ \\
IA827 & 1.747$^c$ \\
NB711 & 2.268$^c$ \\
NB816 & 1.745$^b$ \\
$Y$   & 1.211$^d$ \\
$J$   & 0.871$^d$ \\
$H$   & 0.563$^d$ \\
$K_s$ & 0.364$^d$ \\
\hline
\end{tabular} \\
$^a$ \citet{liske15} \\
$^b$ \citet{capak07} \\
$^c$ \citet{laigle16} \\
$^d$ \citet{mccracken12}
\label{tab:extinct}
\end{table}

\begin{figure*}
\begin{minipage}{5.8in}
\begin{center}
\includegraphics[width=1.0\linewidth]{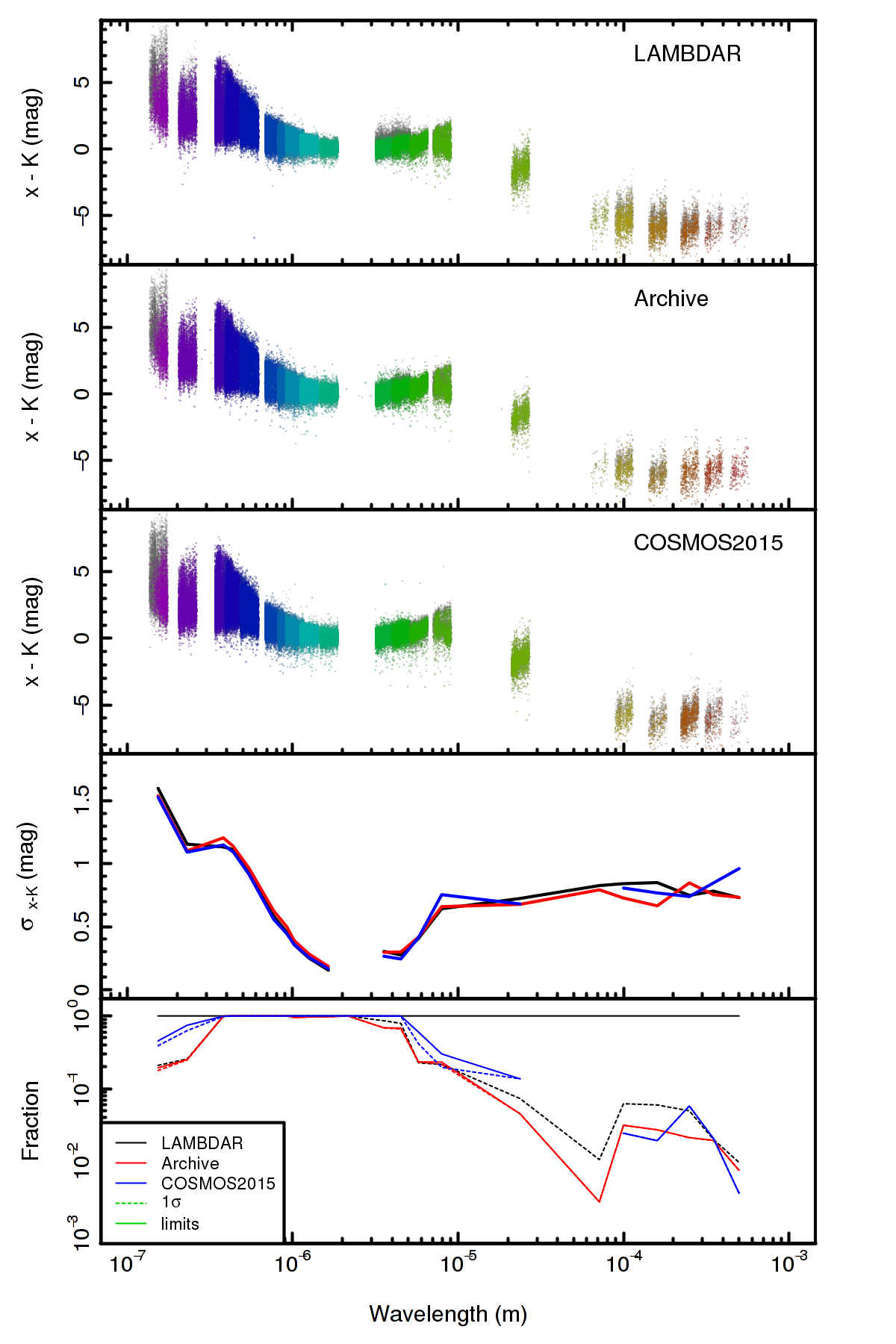}
\caption{Top three panels: colour spreads with respect to $K$ for the indicated data sets. 5$\sigma$ detections are coloured, while 1$\sigma$ detections are in grey. Fourth panel: the spread of the $x-K$ colour distributions, as given by half of the difference between the 84th and 16th percentiles. Bottom panel: fraction of objects with upper limits (solid line) and 1$\sigma$ (dashed line) detections.}
\label{fig:summary}
\end{center}
\end{minipage}
\end{figure*}

We release three sets of catalogues. The main catalogue contains abridged redshift information from \citet{davies15} and fluxes and errors for each band in Janskys. This catalogue contains fluxes corrected for Galactic extinction for $FUV$ to $K_s$ inclusive using the $E(B-V)$ values from the \citet{schlegel98} dust maps multiplied by the $k(\lambda)$ values given in Table \ref{tab:extinct}. Objects which have a greater than 0.8\arcsec positional offset beteen this catalogue and the 2008 COSMOS photometric catalogue have been assigned a FLAGS value of 1. To denote masked regions, we propagate the relevant COSMOS2015 flag column. Non-matches to COSMOS2015 are assigned a mask value equal to their nearest neighbour. For objects that do not have flux measurements due to cascading flux cuts, we assign a flux of -9998 and an error equal to the $1\sigma$ point source limit as calculated from the median sky RMS for 1000 random apertures (see Figure \ref{fig:sky}). The fraction of objects assigned upper limits range from 82 per cent in the ultraviolet, 81 per cent for IRAC 3 and 4, 94 per cent for MIPS 24 and PACS, 96 per cent for SPIRE 250, 98 per cent for SPIRE 350 and 99 per cent for SPIRE 500 and MIPS 70. These are shown by the bottom panel of Figure \ref{fig:summary}.

Figure \ref{fig:summary} attempts to summarise our panchromatic catalogue as compared to COSMOS2015, and what one might construct from existing archival photometry. The upper three panels show, for each catalogue, the colour distribution for the same set of galaxies. The vertical spread therefore indicates a colour range ($x-K$; where $x$ represents each filter in turn), each colour band denotes a filter ($x$), and the horizontal spread the redshift range ($0.4 < z < 0.8$). The density of data points reflects the number of objects. To create the sample the three catalogues were matched and only those galaxies with $f > \delta f$ for both $K$ and $x$ in all three catalogues selected. The fourth panel shows the widths of the colour distribution as measured from the 84th-16th percentile range (i.e. $1\sigma$) which are essentially comparable for all datasets. Finally, the bottom plot shows the fraction of objects with a $1\sigma$ detection (dashed line), or a limit (solid line). From the $u$ to $K$ bands all catalogues contain a measurement for all objects. In the FUV and NUV and IRAC bands a recorded strong or credible measurement occurs more frequently in the COSMOS2015 catalogue, and in the far-IR our catalogue contains more credible measurements than either COSMOS2015 or archival data. However the critical advancement is that our catalogue now contains a measurement or flux limit for every object, essentially increasing the fraction of systems with far-IR constraints from a few per cent to all galaxies in all bands.

Additionally we release the catalogues output by \textsc{lambdar} for each of the 38 bands, which provide more detailed photometric measurements and errors, as separate files. These catalogues also contain warning flags, such as for objects affected by the saturated NaN region replacement detailed in Section \ref{sec:phot2}. The fluxes contained in these catalogues are not extinction corrected.

In addition to the new photometry catalogues presented in the work, we also highlight the release of an updated version of the spectroscopic catalogue outlined in \citet{davies15}. This catalogue now incorporates DR3 of the zCOSMOS \citep{lilly07} spectra, released in January 2015 (ESO Large Programme LP175.A-0839) and the recent photometric redshift analysis of the COSMOS2015 team \citep{laigle16}. Columns 135-139 of the Davies et al. catalogue are updated with the zCOSMOS DR3 measurements and additional columns 176-179 are added for the COSMOS2015 best-fit redshift (`ZP\_COSMOS2015'), the 1$\sigma$ upper and lower error range (`ZL68\_COSMOS2015'  and `ZH68\_COSMOS2015'), and the best fit $\chi^2$ value (`CHI2\_COSMOS2015'). We also update the `Z\_BEST', `Z\_USE' and `Z\_GEN' parameters accordingly. To do this, we undertake a similar process to that outlined in Section 4 of \citet{davies15}, where we compare redshift measurements across various observations. However, we now perform matching in comparison to the zCOSMOS DR3 catalogue instead of the zCOSMOS-bright 10k catalogue, and to the more recent COSMOS2015 photometric redshifts over the original \citet{ilbert09} photometric redshifts. For example, Z\_GEN==5 now refers to a $<10\%$ offset between our \textsc{autoz} measurement and the COSMOS2015 photometric redshift.

This updated catalogue yields an increased number of Z\_USE==1 sources (good high resolution redshift), increasing the sample to $\sim19k$ galaxies, and slightly reducing the spread and outlier rate of comparisons between Z\_BEST and zCOSMOS/PRIMUS/photometric redshifts, as displayed in Figure 9 of \citet{davies15}. We refer the reader to \citet{davies15} for further details of how the spectroscopic catalogue was constructed, and note that the updated catalogue can be found here: \url{http://cutout.icrar.org/G10/G10CosmosCatv03.tar.gz}.

\section{Conclusion}
\label{sec:conclusion}

We have produced a 38 band photometric catalogue in COSMOS spanning from far-ultraviolet to far-infrared wavelengths in a manner consistent with the equivalent \citet{wright16a} GAMA catalogue. We gathered multiwavelength imagery from the \textit{GALEX} Deep Imaging Survey, COSMOS, UltraVISTA, S-COSMOS, SPLASH, PEP and HerMES surveys. From this data, we obtained consistent total flux measurements for a sample of 185,907 sources using \textsc{lambdar}. This required the construction of an aperture catalogue by manually checking and adjusting raw \textsc{SExtractor} output, a process that is prohibitively labour-intensive for the next generation of galaxy surveys. We demonstrate that the resulting photometric catalogue has accurate astrometry, is consistent with existing photometric datasets and achieves adjacent colour distributions --- a proxy for photometric measurement error --- comparable to existing data sets. The released catalogue is complete for objects with $i < 24.5$~mag and partially complete to $i < 25$~mag due to a rigid flux cut made prior to the \textsc{lambdar} measurements. As our catalogue is designed for panchromatic analysis, including SED fitting, we tested it for a sample of 5619 galaxies using \textsc{magphys}. We found improved convergence and goodness of fit with our catalogue compared to table matching archival photometry. The catalogues and a cutout generator for the multiwavelength imagery used are available at \url{http://cutout.icrar.org/G10/dataRelease.php}.

This sample will be used in future observations as an input catalogue for a spectroscopic survey to complete the G10 region. This catalogue will form the basis for a GAMA-equivalent multiwavelength database at intermediate redshifts. This database will enable the derivation of physical properties and structural parameters for the COSMOS region using the same techniques as GAMA and enable comparative studies of the cosmic spectral energy distribution (Andrews et al. in prep), galaxy structure and morphology, star formation rates (Davies et al. in prep), stellar masses (Wright et al. in prep) and panchromatic measurements of the extragalactic background light \citep{driver16b}. In combination with further spectroscopic observations of the G10 region, we will create catalogues of groups (akin to \citealt{robotham11}) and large scale structure. 

In addition to enabling comparisons to low-redshift galaxy evolution surveys, these catalogues will pave the way for future galaxy evolution surveys such as WAVES \citep{waves} and provide a basis for optically motivated stacking using 21~cm data from the COSMOS HI Large Extragalactic Survey (CHILES; \citealt{fernandez13}).

\section*{Acknowledgements}
We wish to thank the anonymous referee whose feedback helped improve this manuscript, the SPLASH team for providing early access to their IRAC images and the COSMOS team for the provision of accurate and precise photometric redshifts.

SKA and AHW are supported by the Australian Government’s Department of Industry Australian Postgraduate Awards (APA). PRK is supported by the Australian Research Council via Discovery Project 140100395.

The G10/COSMOS redshift catalogue and cutout tool uses data acquired as part of the Cosmic Evolution Survey (COSMOS) project and spectra from observations made with ESO Telescopes at the La Silla or Paranal Observatories under programme ID 175.A-0839. The G10 cutout tool is hosted and maintained by funding from the International Centre for Radio Astronomy Research (ICRAR) at the University of Western Australia.

Based on observations obtained with MegaPrime/MegaCam, a joint project of CFHT and CEA/IRFU, at the Canada-France-Hawaii Telescope (CFHT) which is operated by the National Research Council (NRC) of Canada, the Institut National des Science de l'Univers of the Centre National de la Recherche Scientifique (CNRS) of France, and the University of Hawaii. This work is based in part on data products produced at Terapix available at the Canadian Astronomy Data Centre as part of the Canada-France-Hawaii Telescope Legacy Survey, a collaborative project of NRC and CNRS. 

Based on data products from observations made with ESO Telescopes at the La Silla Paranal Observatory under ESO programme ID 179.A-2005 and on data products produced by TERAPIX and the Cambridge Astronomy Survey Unit on behalf of the UltraVISTA consortium.

This work is based in part on observations made with the Spitzer Space Telescope, which is operated by the Jet Propulsion Laboratory (JPL), California Institute of Technology under NASA contract 1407.

This research has made use of data from HerMES project (\url{http://hermes.sussex.ac.uk/}). HerMES is a Herschel Key Programme utilising Guaranteed Time from the SPIRE instrument team, ESAC scientists and a mission scientist. The HerMES data was accessed through the Herschel Database in Marseille (HeDaM - \url{http://hedam.lam.fr}) operated by CeSAM and hosted by the Laboratoire d'Astrophysique de Marseille.

\label{lastpage}


\begin{thebibliography}{99}
\bibitem[\protect\citeauthoryear{Ahn et al.}{2014}]{ahn14} Ahn C.~P. et al., 2014, ApJS, 211, 17 %
\bibitem[\protect\citeauthoryear{Alpaslan et al.}{2014}]{alpaslan14} Alpaslan M. et al. 2014, MNRAS, 438, 177
\bibitem[\protect\citeauthoryear{Aune et al.}{2003}]{aune03} Aune, S. et al. 2003, Proc. SPIE, 4841, 513
\bibitem[\protect\citeauthoryear{Baldry et al.}{2014}]{baldry14} Baldry I.~K. et al. 2014, MNRAS, 441, 2440
\bibitem[\protect\citeauthoryear{Bertin \& Arnouts}{1996}]{bertin96} Bertin E. \& Arnouts S. 1996, A\&A, 117, 393
\bibitem[\protect\citeauthoryear{Boulade et al.}{2003}]{boulade03} Boulade, O. et al. 2003, Proc. SPIE, 4841, 72
\bibitem[\protect\citeauthoryear{Bertin et al.}{2002}]{bertin02} Bertin, E. et al. 2002, Astronomical Data Analysis Software and Systems XI, 281, 228 %
\bibitem[\protect\citeauthoryear{Brammer et al.}{2012}]{brammer12} Brammer G.~B. et al. 2012, ApJS, 200, 13
\bibitem[\protect\citeauthoryear{Capak et al.}{2007a}]{capak07}  Capak P. et al. 2007, ApJS, 172, 99
\bibitem[\protect\citeauthoryear{Capak et al.}{2007b}]{capak07b} Capak P. et al. 2007, ApJS, 172, 284
\bibitem[\protect\citeauthoryear{Coil et al.}{2011}]{coil11} Coil A.~L. et al., 2011, ApJ, 741, 8
\bibitem[\protect\citeauthoryear{Cool et al.}{2013}]{cool13} Cool R.~J. et al., 2013, ApJ, 767, 118
\bibitem[\protect\citeauthoryear{da Cunha et al.}{2008}]{dacunha08} da Cunha E., Charlot S., Elbaz D. 2008, MNRAS 388, 1595
\bibitem[\protect\citeauthoryear{Davies et al.}{2015a}]{davies15} Davies L.~J.~M. et al. 2015a, MNRAS, 447, 1014 %
\bibitem[\protect\citeauthoryear{Davies et al.}{2015b}]{davies15b} Davies L.~J.~M. et al. 2015b, MNRAS, 452, 616
\bibitem[\protect\citeauthoryear{Davies et al.}{2016a}]{davies16a} Davies L.~J.~M. et al. 2016a, MNRAS, in press
\bibitem[\protect\citeauthoryear{Davies et al.}{2016b}]{davies16b} Davies L.~J.~M. et al. 2016b, MNRAS, 455, 4013
\bibitem[\protect\citeauthoryear{de Jong et al.}{2015}]{dejong15} de Jong J.~T.~A. et al. 2015, A\&A, 582, A62
\bibitem[\protect\citeauthoryear{Driver et al.}{2011}]{driver11} Driver S.~P. et al. 2011, MNRAS, 413, 971 %
\bibitem[\protect\citeauthoryear{Driver et al.}{2012}]{driver12} Driver S.~P. et al. 2012, MNRAS, 427, 3244
\bibitem[\protect\citeauthoryear{Driver et al.}{2015}]{waves} Driver S.~P., Davies L.~J., Meyer M., Power C., Robotham A.~S.~G., Baldry I.~K., Liske J., Norberg P. 2015, ASSP, in press (arXiv: 1507.00676) %
\bibitem[\protect\citeauthoryear{Driver et al.}{2016a}]{driver15} Driver S.~P. et al. 2016a, MNRAS, 455, 3911 %
\bibitem[\protect\citeauthoryear{Driver et al.}{2016b}]{driver16b} Driver S.~P. et al. 2016b, ApJ, in press
\bibitem[\protect\citeauthoryear{Eales et al.}{2010}]{eales10} Eales S. et al., 2010, PASP, 122, 499 %
\bibitem[\protect\citeauthoryear{Edge et al.}{2013}]{edge13} Edge A., Sutherland W., Kuijken K., Driver S.~P., McMahon R., Eales S., Emerson J.~P. 2013, The Messenger, 154, 32 %
\bibitem[\protect\citeauthoryear{Fern\'{a}ndez et al.}{2013}]{fernandez13} Fern\'{a}ndez X et al. 2013, ApJ, 770, L29
\bibitem[\protect\citeauthoryear{Frayer et al.}{2009}]{frayer09} Frayer D.~T. et al. 2009, AJ, 138, 1261
\bibitem[\protect\citeauthoryear{Garilli et al.}{2008}]{garilli08} Garilli B. et al., 2008, A\&A, 486, 683
\bibitem[\protect\citeauthoryear{Gordon et al.}{2008}]{gordon08} Gordon K.~D., Engelbracht C.~W., Rieke G.~H., Misselt K.~A., Smith J.-D.~T., Kennicutt R.~C. 2008, ApJ, 682, 336 %
\bibitem[\protect\citeauthoryear{Griffin et al.}{2010}]{griffin10} Griffin M. J. et al. 2010, A\&A, 518, L3
\bibitem[\protect\citeauthoryear{Hill et al.}{2011}]{hill11} Hill D.~T. et al. 2011, MNRAS, 412, 765
\bibitem[\protect\citeauthoryear{Hora et al.}{2012}]{hora12} Hora J.~L. et al. 2012, Proc. SPIE, 8442, 39
\bibitem[\protect\citeauthoryear{Hsieh et al.}{2012}]{hsieh12} Hsieh B.-C., Wang W.-H., Hsieh C.-C., Lin L., Yan H., Lim J., Ho, P.~T.~P. 2012, ApJS, 203, 23
\bibitem[\protect\citeauthoryear{Ilbert et al.}{2009}]{ilbert09} Ilbert O. et al. 2009, ApJ, 690, 1236
\bibitem[\protect\citeauthoryear{Ilbert et al.}{2015}]{ilbert15} Ilbert O. et al. 2015, A\&A, 579, A2
\bibitem[\protect\citeauthoryear{Kelvin et al.}{2012}]{kelvin12} Kelvin L.~S. et al. 2012, MNRAS, 421, 1007
\bibitem[\protect\citeauthoryear{Komiyama et al.}{2003}]{komiyama03} Komiyama, Y. et al. 2003, Proc. SPIE, 4841, 152
\bibitem[\protect\citeauthoryear{Laidler et al.}{2007}]{laidler07} Laidler V.~G. et al. 2007, PASP, 119, 1325
\bibitem[\protect\citeauthoryear{Laigle et al.}{2016}]{laigle16} Laigle C. et al. 2016, ApJS, 224, 24
\bibitem[\protect\citeauthoryear{Lawrence et al.}{2007}]{lawrence07} Lawrence A. et al. 2007, MNRAS, 379, 1599
\bibitem[\protect\citeauthoryear{Le F\'{e}vre et al.}{2015}]{lefevre15} Le F\'{e}vre O. et al. 2015, A\&A, 576, A79
\bibitem[\protect\citeauthoryear{Le Floc'h et al.}{2009}]{lefloch09} Le Floc'h E. et al. 2009, ApJ, 703, 222
\bibitem[\protect\citeauthoryear{Levenson et al.}{2010}]{levenson10} Levenson L. et al. 2010, MNRAS, 409, 83
\bibitem[\protect\citeauthoryear{Lilly et al.}{2007}]{lilly07} Lilly S.~J. et al. 2007, ApJS, 172, 70
\bibitem[\protect\citeauthoryear{Lilly et al.}{2009}]{lilly09} Lilly S.~J. et al. 2009, ApJS, 184, 218
\bibitem[\protect\citeauthoryear{Liske et al.}{2015}]{liske15} Liske J. et al. 2015, MNRAS, 452, 2087 %
\bibitem[\protect\citeauthoryear{Loveday et al.}{2015}]{loveday15} Loveday J. et al. 2015, MNRAS, 451, 1540
\bibitem[\protect\citeauthoryear{Lutz et al.}{2011}]{lutz11} Lutz D. et al. 2011, A\&A, 532, A90
\bibitem[\protect\citeauthoryear{Martin et al.}{2005}]{martin05} Martin C. et al., 2005, ApJ, 619, 1
\bibitem[\protect\citeauthoryear{Masters et al.}{2012}]{masters12} Masters D. et al. 2012, ApJ, 755, 2
\bibitem[\protect\citeauthoryear{McCracken et al.}{2012}]{mccracken12} McCracken H.~J. et al. 2012, A\&A, 544, A156
\bibitem[\protect\citeauthoryear{Moffett et al.}{2016}]{moffett16} Moffett A. et al. 2016, MNRAS, 457 1308
\bibitem[\protect\citeauthoryear{Muzzin et al.}{2013}]{muzzin14} Muzzin A. et al. 2013, ApJS, 206, 8 %
\bibitem[\protect\citeauthoryear{Oliver et al.}{2012}]{oliver12} Oliver S.~J. et al. 2012, MNRAS, 424, 1614 %
\bibitem[\protect\citeauthoryear{Pilbratt et al.}{2010}]{pilbratt10} Pilbratt G.~L. et al. 2010, A\&A, 518, L1 %
\bibitem[\protect\citeauthoryear{Poglitsch et al.}{2010}]{poglitsch10} Poglitsch A. et al. 2010, A\&A, 518, L2 %
\bibitem[\protect\citeauthoryear{Riguccini et al.}{2015}]{riguccini15} Riguccini L. et al. 2015, MNRAS, 452, 470
\bibitem[\protect\citeauthoryear{Robotham et al.}{2011}]{robotham11} Robotham A.~S.~G. et al. 2011, MNRAS, 416, 2640
\bibitem[\protect\citeauthoryear{Roseboom et al.}{2010}]{roseboom10} Roseboom I.~G. et al. 2010, MNRAS, 409, 48
\bibitem[\protect\citeauthoryear{Sanders et al.}{2007}]{sanders07} Sanders D.~B. et al. 2007, ApJS, 172, 86
\bibitem[\protect\citeauthoryear{Schlegel, Finkbeiner \& Davis}{1998}]{schlegel98} Schlegel D.~J., Finkbeiner D.~P., Davis M., 1998, ApJ, 500, 525 %
\bibitem[\protect\citeauthoryear{Scoville et al.}{2007}]{scoville07} Scoville N. et al. 2007, ApJS, 172, 1
\bibitem[\protect\citeauthoryear{Silverman et al.}{2015}]{silverman15} Silverman J. D. et al. 2015, ApJS, 220, 12
\bibitem[\protect\citeauthoryear{Smith et al.}{2012}]{smith12} Smith A.~J. et al. 2012, MNRAS, 419, 377
\bibitem[\protect\citeauthoryear{Taniguchi et al.}{2007}]{taniguchi07} Taniguchi Y. et al. 2007, ApJS, 192, 9
\bibitem[\protect\citeauthoryear{Taniguchi et al.}{2015}]{taniguchi16} Taniguchi Y. et al. 2015, PASJ, 67, 104 %
\bibitem[\protect\citeauthoryear{Taylor et al.}{2011}]{taylor11} Taylor E.~N. et al. 2011, MNRAS, 418, 1587
\bibitem[\protect\citeauthoryear{Viero et al.}{2013}]{viero13} Viero M.~P. et al. 2013, ApJ, 772, 77
\bibitem[\protect\citeauthoryear{Wang et al.}{2014}]{wang14} Wang L. et al. 2014, MNRAS, 444, 2870
\bibitem[\protect\citeauthoryear{Wright et al.}{2010}]{wright10} Wright E.~L. et al., 2010, AJ, 140, 1868 %
\bibitem[\protect\citeauthoryear{Wright et al.}{2016}]{wright16a} Wright A.~H. et al. 2016, MNRAS, 460, 765
\bibitem[\protect\citeauthoryear{York et al.}{2000}]{york00} York D., et al., 2000, AJ, 120, 1579 %
\bibitem[\protect\citeauthoryear{Zamojski et al.}{2007}]{zamojski07} Zamojski M.~A. et al. 2007, ApJ, 172, 468

\end{thebibliography}
\end{document}